\documentclass[a4paper,pre,twocolumn,amsmath,amssymb,longbibliography,nofootinbib]{revtex4-1}
\usepackage{bm}
\usepackage{graphicx}
\usepackage{xcolor}
\usepackage{soul}
\usepackage[colorlinks,linkcolor=magenta,citecolor=magenta]{hyperref}
\usepackage{multirow}
\usepackage[capitalise]{cleveref}

\newcommand{\dd}{\mathrm{d}}

\newcommand{\fd}[2]{\frac{\delta #1}{\delta #2}}
\newcommand{\mean}[1]{\langle #1 \rangle}
\newcommand{\Int}[1]{\int\dd #1\;}
\newcommand{\IInt}[3]{\int_{#2}^{#3}\dd #1\;}
\renewcommand{\vec}[1]{\mathbf #1}

\newcommand{\al}{\alpha}
\newcommand{\gam}{\gamma}
\newcommand{\Gam}{\Gamma}
\newcommand{\eps}{\varepsilon}
\newcommand{\kap}{\kappa}
\newcommand{\lam}{\lambda}
\newcommand{\Lam}{\Lambda}

\newcommand{\om}{\omega}
\newcommand{\Om}{\Omega}

\newcommand{\x}{\vec r}

\newcommand{\im}{\text{i}}

\AtBeginDocument{%
    \newwrite\bibnotes
    \def\bibnotesext{Notes.bib}
    \immediate\openout\bibnotes=\jobname\bibnotesext
    \immediate\write\bibnotes{@CONTROL{REVTEX41Control}}
    \immediate\write\bibnotes{@CONTROL{%
    apsrev41Control,author="08",editor="1",pages="1",title="0",year="1"}}
     \if@filesw
     \immediate\write\@auxout{\string\citation{apsrev41Control}}%
    \fi
}%

\begin{document}

\title{Dynamic renormalization of scalar active field theories}

\author{Nikos Papanikolaou}
\author{Thomas Speck}
\affiliation{Institute for Theoretical Physics IV, University of Stuttgart, Heisenbergstr. 3, 70569 Stuttgart, Germany}


\begin{abstract}
  We study Active Model B+, a scalar field theory extending the paradigmatic Model B for equilibrium coexistence through including terms that do not arise from an underlying free energy functional and thus break detailed balance. In the first part of the manuscript, we provide a pedagogical and self-contained introduction to one-loop dynamic renormalization. We then address the technical challenge of complex vertex functions through developing a symbolic computer algebra code that allows us to obtain the graphical corrections of model parameters. We argue that the additional terms of Active Model B+ imply the generation of, potentially relevant, higher-order terms; strongly restricting the parameter regime in which we can apply a perturbative renormalization scheme. Moreover, we elucidate the role of the cubic coefficient, which, in contrast to passive Model B, is incessantly generated by the new terms. Analyzing its behavior with and without field shift near the Wilson-Fisher fixed point, we find that additional fixed points in the one-loop flow equations are likely artifacts. Additionally, we characterize the renormalization flow of perturbatively accessible field theories derived from Active Model B+.
\end{abstract}

\maketitle


\section{Introduction}

In statistical physics, we often encounter dynamic (or ``kinetic'') equations of the type
\begin{equation}
  \partial_t\phi = F(\phi,\nabla\phi,\dots;\vec x) + \eta
  \label{eq:phi:F}
\end{equation}
describing the stochastic evolution of a real-valued scalar field $\phi(\x,t)$ in $d$ dimensions ($\x\in\mathcal D\subseteq\mathbb R^d$). This could be the density change of a monocomponent fluid, related to the composition of a binary mixture, or the (relative) height of a film covering a surface. The details of the specific model under scrutiny are encoded in the function $F(\phi,\nabla\phi,\dots;\vec x)$ of the field and its derivatives, which we assume can be parametrized by a set of model parameters $\vec x=(x_1,\dots)$. Since $\phi(\x,t)$ is already a coarse-grained description of the microscopic degrees of freedom it is accompanied by a noise field $\eta(\x,t)$.

We might want to quantify the behavior of Eq.~\eqref{eq:phi:F} at large scales, at which many details of the microscopic interactions are irrelevant. A paradigmatic example is the (noisy) Navier-Stokes equation (for the vectorial velocity field), which is characterized by a single material parameter, the viscosity, and constrained by conservation and symmetry laws. A systematic tool is renormalization~\cite{goldenfeld18}, which proceeds through the step-wise coarse-graining of the field $\phi(\x,t)$ through integrating small-scale features, relating the reduction of degrees of freedom to changes of the model parameters $\vec x(\ell)$ as a function of a scale variable $\ell$. Based on their scaling dimensions, the relevant model parameters can be identified. Renormalization in statistical physics is of course closely connected to the understanding of critical phenomena, which have posed a major theoretical challenge due to the crucial role of fluctuations caused by a diverging correlation length.

One flavor of renormalization is \emph{dynamic renormalization}, which operates directly on the level of the evolution equation~\eqref{eq:phi:F}. Originally developed in the context of ferromagnets~\cite{ma75} and turbulence~\cite{forster77}, it is well suited for non-equilibrium systems lacking a free energy functional and has been applied to a wide range of systems out of equilibrium such as surface growth models~\cite{kardar86,sun89,haselwandter08,zhang08}, polymers in random media~\cite{medina89}, reaction-diffusion models~\cite{gagnon15}, population dynamics~\cite{zinati22}, chemotaxis of bacteria~\cite{gelimson15,mahdisoltani21}, and more~\cite{ertas93,basu09,dolai17}. More recently, non-equilibrium systems in which detailed balance is broken on the level of individual constituents, often summarized as ``active matter'', have moved into the focus. For example, active fluids composed of millions of self-propelled or self-spinning particles can now be realized in experiment~\cite{soni19,chardac21}, displaying a wealth of collective phenomena. On the modeling side, great effort has been invested into deriving continuum models that capture the collective behavior and allow to isolate the underlying mechanisms. A number of these continuum descriptions have been studied through renormalization techniques: polar alignment into flocks and swarms~\cite{toner98,cavagna19,skultety20,cavagna23,chen24,miller24}, coupled to birth and death processes~\cite{chen20}, active nematics~\cite{mishra10,shankar18a}, active membranes~\cite{cagnetta22,cagnetta22a}, and non-aligning active particles~\cite{caballero18,maggi22}.

Active matter systems are sometimes classified by the degree of the emerging order: nematic, polar, and scalar. The latter can thus be described by a single scalar field, typically related to density, and among others applies to fluids of \emph{non-aligning} self-propelled particles. The interplay of directed motion with volume exclusion gives rise to motility-induced phase separation~\cite{buttinoni13,cates15}, a dynamic instability in which the effective propulsion speed is reduced as the local density increases. Even in the absence of cohesive forces, coexistence of dense domains with an active gas is observed. This coexistence terminates in a critical point, which has been reported to fall into the same Ising universality as passive fluids governed by attractive forces~\cite{maggi21,partridge19}, with Refs.~\cite{siebert18,dittrich21} reporting deviations from the expected Ising critical exponents in two dimensions. Motility-induced phase separation superficially resembles passive phase separation captured by the continuum ``Model B'' (which stems from the famous classification of Hohenberg and Halperin~\cite{hohenberg77}, see Table~I therein). This is the premise for Active Model B+, which adds the lowest-order derivatives of the field that cannot be derived as a functional derivative of a free energy~\cite{tjhung18}. One-loop dynamic renormalization has been applied to Active Model B+ by Caballero and Cates~\cite{caballero18}, reporting two relevant fixed points in addition to the Wilson-Fisher fixed point~\cite{wilson72}. Their analysis left out several graphs [cf. Fig.~\ref{fig:graphs}] and was based on an invariant ratio of model parameters that we cannot confirm. More importantly, however, we will argue that Active Model B+ is ``incomplete'', which necessitates an expansion of vertex functions in model parameters in addition to the $\eps$-expansion.

Here we revisit Active Model B+ from the perspective of dynamic renormalization. Alternatively, stochastic dynamic equations [such as Eq.~\eqref{eq:phi:F}] can be transformed into a stochastic action (known as Doi-Peliti~\cite{doi76,peliti85} and Martin-Siggia-Rose-DeDominicis-Janssen~\cite{martin73,janssen76,weber17} formalisms), for which methods from (quantum) field theory are more directly applicable~\cite{tauber07}. It has been applied successfully in particular to systems with discrete ``chemical'' events such as reaction-diffusion models~\cite{tauber05a,cardy98,tauber12a,winkler12}, percolation~\cite{janssen05}, and the voter model~\cite{janssen05a,garcia-millan20}. However, the field involved in this action no longer corresponds to the physical field $\phi$ and alternative formulations loose the simple structure that makes the action amenable for perturbative renormalization~\cite{lefevre07}. Moreover, the ``graphical language'' needed for bookkeeping terms is more complex than for dynamic renormalization.

Our manuscript is organized as follows. In Sec.~\ref{sec:method}, we give a brief and concise introduction to dynamic renormalization and directed graphs as a tool to represent integrals. The purpose of this section is to expose the minimal ``machinery'' to obtain one-loop flow equations and to study their properties. By no means is it a replacement for more detailed reviews~\cite{hohenberg77}, lecture notes~\cite{tauber07,tauber12}, and books~\cite{goldenfeld18,tauber13} on renormalization. In Sec.~\ref{sec:ambp}, we then revisit Active Model B+ and find new non-linear terms (Secs.~\ref{sec:cubic} and \ref{sec:higher}). We then describe two possible ways to analyze the resulting flow equations (Sec.~\ref{sec:ambp_flow}). Finally, in Sec.~\ref{sec:models} we apply dynamic renormalization to two scalar field theories: a surface growth field theory which, to the best of our knowledge, has not been studied before (Sec.~\ref{sec:mod_ckpz}) and a neural network model (Sec.~\ref{sec:neural}).


\section{Dynamic renormalization}
\label{sec:method}

\subsection{Linear theory}
\label{sec:basics}

As for any perturbative approach, we need a reference around which we expand exploiting small parameters. In the case of dynamic renormalization, this reference is the linear theory obtained through neglecting non-linear terms with $F_0=-(\im\nabla)^\al(a\phi-\kap\nabla^2\phi)$. We switch to Fourier space through
\begin{equation}
  \phi(\x,t) = \int\frac{\dd\om}{2\pi}\int\frac{\dd^d \vec q}{(2\pi)^d}e^{\im\vec q\cdot\x-\im\om t} \phi(\om,\vec q),
\end{equation}
where we use the same symbol for the field but with different arguments. The linearized evolution equation~\eqref{eq:phi:F} then reads
\begin{equation}
  -\im\om\phi = -q^\al(a+\kap q^2)\phi + \eta
  \label{eq:phi:lin}
\end{equation}
with solution $\phi(\om,\vec q)=G_0(\om,q)\eta(\om,\vec q)$, where we have defined the bare propagator
\begin{equation}
  G_0(\om,q) \equiv \frac{1}{-\im\om+q^\al(a+\kap q^2)} = \frac{1}{-\im\om+h(q)}
  \label{eq:G0}
\end{equation}
with $h(q)\equiv q^\al(a+\kap q^2)$ for later use. Throughout, we will write $q$ for the magnitude of the wave vector $\vec q$ (and analogously for other wave vectors). For the noise correlations, we will employ
\begin{equation}
  K(\hat q,\hat q') \equiv \mean{\eta(\hat q)\eta(\hat q')} \\ = 2Dq^\al(2\pi)^{d+1}\delta^d(\vec q+\vec q')\delta(\om+\om'),
  \label{eq:K}
\end{equation}
where, for ease of notation, we have combined frequency $\om$ and wave vector $\vec q$ into the single vector $\hat q\equiv(\om,\vec q)$ with $d+1$ components. Note that for a conserved field the factor $q^2$ suppresses fluctuations of the integrated field as required. The strength of the noise is quantified by the coefficient $D$. For dynamics obeying detailed balance, the fluctuation-dissipation theorem constraints $D$ to be related to the temperature but in the following we will mostly treat it as another free model parameter.

It is now straightforward to calculate the field correlations
\begin{equation}
  \mean{\phi(\hat q)\phi(\hat q')} = (2\pi)^{d+1}C_0(\hat q)\delta^{d+1}(\hat q+\hat q')
  \label{eq:corr}
\end{equation}
with the dynamic structure factor
\begin{equation}
  C_0(\om,q) \equiv 2Dq^\al G_0(\om,q)G_0(-\om,q) = \frac{2Dq^\al}{\om^2+[h(q)]^2}.
  \label{eq:C0}
\end{equation}
The static structure factor follows immediately as
\begin{equation}
  S_0(q) \equiv \int_{-\infty}^\infty\frac{\dd\om}{2\pi}C_0(\om,q) = \frac{D}{a+\kap q^2}
  \label{eq:S0}
\end{equation}
independent of $\al$. Clearly, we can construct one length scale, $\xi=(a/\kap)^{-1/2}$, which is the correlation length governing the exponential decay of correlations in real space. With this correlation length, the static structure factor becomes
\begin{equation}
  S_0(q) = \frac{(D/\kap)\xi^2}{1+(\xi q)^2}.
  \label{eq:S0:xi}
\end{equation}
For $a\to 0$ the correlation length $\xi\to\infty$ diverges with $S_0(q)=(D/\kap)q^{-2}$.

\subsection{Basic idea of renormalization}
\label{sec:rg}

Our model is useful down to a length scale $\Lam^{-1}$ (typically related to the particle size or the lattice spacing) below which we have no further information. In Fourier space this implies that $\phi(\om,q\geqslant\Lam)=0$. Let us collect the model parameters into the vector $\vec x=\vec x(\Lam)$ depending on the microscopic cut-off. Now we integrate out spatial features on length scales smaller than $b\Lam^{-1}$ with a factor $b>1$. We thus lose microscopic information (corresponding to large $q\sim\Lam$) and consequently will need new parameters $\vec x\mapsto\vec x'=\vec x(\Lam/b)$ to describe the evolution of the field.

\begin{figure}[t]
  \centering
  \includegraphics{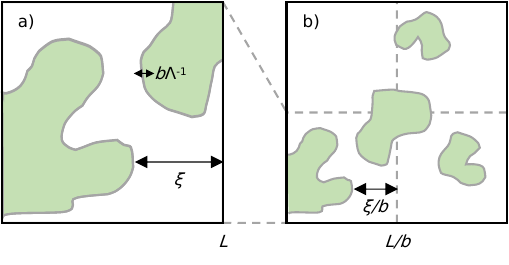}
  \caption{Sketch of the renormalization procedure in real space. (a)~We start with a system of length $L$, microscopic cut-off $\Lam^{-1}$, and correlation length $\xi$. After integrating out degrees of freedom on the smallest scales the cut-off length is increased to $b\Lam^{-1}$. (b)~We now zoom out by the factor $b$ to look at a larger portion of the system with restored cut-off $\Lam^{-1}$ and reduced correlation length $\xi/b$.}
  \label{fig:renorm}
\end{figure}

Before we repeat this step to get to an even coarser scale, let us ``zoom out'' by the factor $b$ and look at a larger portion of our system (Fig.~\ref{fig:renorm}). Holding the absolute size $L$ fixed, this means that coordinates $\x$ in our original system have shrunk to $\x/b$. Moreover, the size of any ``structure'' has shrunk by the same factor, in particular the correlation length $\xi\mapsto\xi'=\xi/b$. Note that this step restores the size of the smallest discernible features to $\Lam^{-1}$ (with respect to $L$).\footnote{As an analogy, consider looking at the system through a microscope at maximal magnification. You then reduce the magnification by a factor $b$ looking at a larger portion but with fixed field of view $L$ and fixed resolution $\Lam^{-1}$ (say, the size of a pixel). An excellent visualization for the Ising model can be found here: \url{https://www.youtube.com/watch?v=MxRddFrEnPc}.} Repeating this procedure induces an evolution, a ``flow'' in model space, during which we zoom out further and further. For an infinitesimal step $b=1+\delta\ell$ we find
\begin{equation}
  x'_i = x_i(\Lam/b) = x_i(\Lam) + \beta_i(\vec x)\delta\ell + \mathcal O(\delta\ell^2)
\end{equation}
with flow equations $\partial_\ell x_i=\beta_i(\vec x)$ implying the solution $\vec x(\ell)$ of model parameters as a function of $\ell$. Keeping explicitly track of the scale $b(\ell+\delta\ell)=b(\ell)(1+\delta\ell)$ yields $b=e^\ell$ and thus $\Lam(\ell)=\Lam_0e^{-\ell}$ is the actual value of the cut-off with initial value $\Lam_0$.

The remaining task is to find the functions $\beta_i$. Of particular interest are fixed points $\beta_i(\vec x^\ast)=0$ and the evolution around these fixed points. Their importance can be appreciated by noting that any initial correlation length $0<\xi_0<\infty$ will flow to $\xi(\ell)=\xi_0e^{-\ell}\to0$ except for points in our model space where the correlation length $\xi\to\infty$ diverges, which will be mapped to \emph{critical} fixed points.

\begin{figure*}
  \centering
  \includegraphics{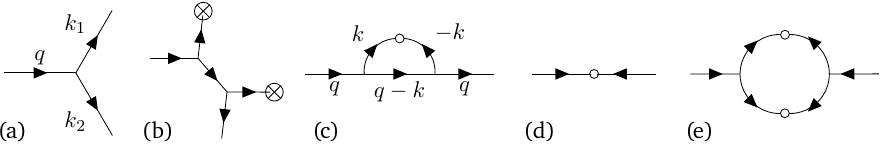}
  \caption{Constructing a graph. (a)~Initial bare vertex with $n=2$ fields as outgoing lines. (b)~Here we replace the upper line by the linear solution $\phi^+$ (indicated with a crossed dot) and attach a new vertex to the lower line. In the second step, we replace one of the two fields by $\phi^+$. (c)~We now join the two $\phi^+$ lines so that their wave vectors cancel, which completes this graph. The last step is to label all lines with their wave vectors. There are $2\times2=4$ ways to arrive at this graph. (d)~Correlation function $C_0$ and (e)~its first graphical correction due to two $2$-vertices.}
  \label{fig:graph}
\end{figure*}

\subsection{Scaling dimensions}
\label{sec:scale}

The change $x\mapsto b^{\Delta_x}x$ of a quantity $x$ under this rescaling procedure defines its \emph{scaling dimension} $\Delta_x$. If a scaling dimension $\Delta_x<0$ is negative then it is called \emph{irrelevant}: Going to larger scales (large $b$) the influence of $x$ is diminished and eventually vanishes. Correspondingly, $\Delta_x>0$ is called \emph{relevant} and the borderline $\Delta_x=0$ \emph{marginal}. Of particular importance is the Gaussian fixed point (G) at which \emph{all} non-linear terms become irrelevant and we are left with the linear theory introduced in Sec.~\ref{sec:basics}.

We already know the scaling dimension $\Delta_\xi=-1$ of the correlation length. For the model parameters, we can derive a number of relations through constraining the form of the evolution equation to remain invariant. Implementing the rescaling step amounts to defining new wave vectors $q'=bq$ together with $\om'=b^z\om$, where $z$ is the dynamical exponent. Inserting both rescaled quantities into the bare propagator [Eq.~\eqref{eq:G0}] yields
\begin{equation}
  G_0(\om'/b^z,q'/b;\kap,a) = b^zG_0(\om',q';\kap'=b^{\Delta_\kap}\kap,a'=b^{\Delta_a}a)
\end{equation}
with $\Delta_\kap=z-\al-2$ and $\Delta_a=z-\al$. While the functional form of $G_0$ is the same, the model parameters have changed. Clearly, the correlation length $\xi'=(\kap'/a')^{1/2}=\xi/b$ then indeed transforms as a length. Demanding that the dynamic equation~\eqref{eq:phi:F} is invariant leads to another relation: the left side acquires a factor $b^{z+\Delta_\phi}$ and, therefore, the noise term scales with $b^{(\Delta_D+\al+d+z)/2}$ [cf. Eq.~\eqref{eq:K}] and thus $\Delta_D=2\Delta_\phi-d+z-\al$.

To get a different view on scaling we briefly return to real space. The Fourier transform of $S_0(q)$ [Eq.~\eqref{eq:S0}] yields the static correlations
\begin{equation}
  \mean{\phi(\x)\phi(\x')} \sim \frac{1}{|\x-\x'|^{d-2}}
  \label{eq:phi:corr}
\end{equation}
for $|\x-\x'|\ll\xi$ ignoring numerical factors. If we demand that the functional form of these correlations does not change--we say that they are \emph{scale invariant}--then a rescaling $\x\mapsto\x/b$ of lengths on the right-hand side has to be compensated by a rescaling of the field, $\phi\mapsto b^{\Delta_\phi}\phi$, with ``naive'' scaling dimension $\Delta^0_\phi=(d-2)/2$. In general, the scaling dimension of the field is
\begin{equation}
  \Delta_\phi = \frac{d-2+\eta}{2}
  \label{eq:sd:phi}
\end{equation}
with $\eta$ known as the \emph{anomalous dimension} (not to be confused with the noise field). Assuming that $D$ is constant (as in thermal equilibrium, $\Delta_D=0$) yields the known relation $z=2+\al-\eta$~\cite{hohenberg77}. Plugging in the above relations, we see that $\eta=\Delta_D-\Delta_\kap$ stems from the mismatch between the scaling dimensions of noise strength $D$ and $\kap$.

From the scaling dimensions, we can extract critical exponents for measurable observables. One scenario is that we need to tune a control parameter $\tau$ (typically the reduced temperature) to a specific value ($\tau=0$). The correlation length then diverges as $\xi\sim|\tau|^{-\nu}$ approaching the critical point, whereby the exponent $\nu=1/\Delta_\tau$ follows from $\Delta_\xi=-1$. The order parameter behaves as $\mean{\phi}\sim(-\tau)^\beta$ with exponent $\beta=\nu\Delta_\phi$. And finally, the susceptibility $S_0(q\to 0)\sim|\tau|^{-\gam}\sim\xi^{2-\eta}$ diverges with exponent $\gam=\nu(2-\eta)$, cf. Eq.~\eqref{eq:S0:xi}.


\subsection{Perturbation series and vertex functions}

We now include non-linear terms into the evolution equation~\eqref{eq:phi:lin}. The $n$-th power of the field in Fourier space becomes
\begin{multline}
  [\phi(\x,t)]^n \to [\phi(\hat k_1)\cdots\phi(\hat k_n)]_{\hat q} = [\phi^n]_{\hat q} \\ = \left[\prod_{i=1}^n\int_{\hat k_i}\phi(\hat k_i)\right](2\pi)^{d+1}\delta(\hat q-\sum_{i=1}^n \hat k_i)
  \label{eq:phi:n}
\end{multline}
with integral
\begin{equation}
  \int_{\hat k} \equiv \int_{|\vec k|<\Lam}\frac{\dd\Om\dd^d \vec k}{(2\pi)^{d+1}}
\end{equation}
and $\hat k\equiv(\Om,\vec k)$. Using this bracket notation, the evolution equation for the field reads
\begin{equation}
  \phi(\hat q) = G_0(\hat q)\eta(\hat q) + G_0(\hat q)\sum_{n=2,\dots}[v_n\phi^n]_{\hat q}
  \label{eq:phi}
\end{equation}
with vertex functions $v_n(\vec k_1,\dots,\vec k_n|\vec q)$. The dependence on wave vectors arises through spatial derivatives of the field in real space. The rule is to replace $\nabla\to-\im\vec q$ for $\nabla$'s acting on everything to their right and $\nabla\to\im\vec k_i$ inside brackets. For example, $\nabla^2\phi^2\to-q^2\phi(\hat k_1)\phi(\hat k_2)$ while $|\nabla\phi|^2\to-\vec k_1\cdot\vec k_2\phi(\hat k_1)\phi(\hat k_2)$. The vertex functions have to be symmetric with respect to exchanging wave vectors $\vec k_i$ since we have only one field.

Clearly, Eq.~\eqref{eq:phi} is not closed since it contains $\phi$ on both sides of the equation. Nevertheless, it can be used to generate the solution as a series of terms with increasing powers of the vertex strengths through inserting into itself. Assuming that these strengths are small implies a perturbation approach close to the Gaussian fixed point. The problem becomes clear immediately: relevant non-linear strengths grow under rescaling, taking them away from the region where the perturbative solution is valid. Our hope, thus, is to discover new fixed points in the vicinity of the Gaussian fixed point and to study their properties.

For consistency, it is helpful to define $v_1(q)\equiv G_0(0,q)=1/h(q)$ for the propagator and $v_0(q)\equiv C_0(0,q)$ for the correlation function. Here we have already set $\om=0$ since a Taylor expansion with respect to $\om$ simply generates derivatives with respect to $t$ in the time representation, which are absent in the original evolution equation~\eqref{eq:phi:F}. Inserting the solution Eq.~\eqref{eq:phi}, the goal is to determine how the vertex functions change after integrating out small-scale features, $[v_n\phi^n]_{\hat q}\to[\tilde v_n\phi^n]_{\hat q}$, with the change of $v_0\to\tilde v_0$ and $v_1\to\tilde v_1$ given through Eq.~\eqref{eq:corr} and Eq.~\eqref{eq:phi}, respectively.

\subsection{Constructing graphs}

To keep track of the terms contributing to the solution it is helpful to employ a graphical language that is inspired by scalar Feynman diagrams (although it lacks the interpretation as particles and momenta)~\cite{weinzierl22}. Each term of the perturbation series is represented as a directed graph $\Gam_n$. We only need very few graphical elements: external lines, internal lines connecting vertices, and a ``sink'' of internal lines representing the correlation function $C_0$. Each vertex has one incoming line and $n$ outgoing lines [e.g., Fig.~\ref{fig:graph}(a) for $n=2$]. A single ($n=1$) outgoing line represents the propagator while the correlation function [Fig.~\ref{fig:graph}(d,e)] has only incoming lines ($n=0$). The sum of outgoing wave vectors $\sum_{i=1}^n\hat k_i=\hat q$ of each vertex equals the incoming wave vector $\hat q$, which is enforced by the $\delta$-distribution in Eq.~\eqref{eq:phi:n}.

To construct the final graph $\Gam_n$ from a bare initial vertex we iteratively either [Fig.~\ref{fig:graph}(b)]
\begin{itemize}
  \item replace a line by the linear solution $\phi\to\phi^+=G_0\eta$ or
  \item attach a vertex to one outgoing line (this line becomes the internal incoming line of the new vertex).
\end{itemize}
Finally, all intermediate $\phi^+$ lines need to end in an open dot, which joins exactly two lines so that the sum of their wave vectors vanishes [Fig.~\ref{fig:graph}(c)]. This step implicitly performs the average over the noise and the open dot plus the two lines together represent the correlation function $C_0(\hat k)$ [Fig.~\ref{fig:graph}(d)]. It should be easy to see that there are multiple ways to arrive at the same final graph $\Gam_n$. Section~\ref{sec:mult} shows how to calculate the multiplicity $|\Gam_n|$ as the number of permutations in the construction of the graph. For example, the multiplicity of the graph Fig.~\ref{fig:graph}(c) is $|\Gam_1|=2\times2=4$ because for each of the two steps there are two possibilities.

\subsection{From graph to integral}

The final graph can then be translated into one or several nested integrals $\mathcal I(\Gam_n;\vec x)$. All internal lines connecting two vertices represent $G_0(\hat k)$ with the corresponding wave vector. All external outgoing lines represent fields $\phi$ with one exception: If there is exactly one outgoing line ($n=1$) then its wave vector is necessarily $\hat q$ and it also represents $G_0(\hat q)$. In the following, for the final graphs we use the convention that the single incoming wave vector is $\vec q$, outgoing wave vectors are $\vec p_i$ with $\sum_i\vec p_i=\vec q$, and internal wave vectors are $\vec k_i$ that will be integrated out. Each internal wave vector necessarily implies a corresponding loop in the graph. For example, reading the graph Fig.~\ref{fig:graph}(c) from left to right leads to
\begin{multline}
  \mathcal I(\Gam_1;\vec x) = 4\int_{\hat k}
  G_0(\hat q)v_2(\vec k,\vec q-\vec k)C_0(\hat k) \\ \times G_0(\hat q-\hat k)v_2(\vec q,-\vec k)G_0(\hat q)
  \label{eq:int:v2}
\end{multline}
and we need to integrate out $\hat k$ to obtain the lowest-order correction to the bare propagator $G_0(0,q)$. The pre-factor is the multiplicity of the graph.

Summing over all distinct graphs with the same number $n$ of outgoing lines yields the ``graphical corrections''
\begin{multline}
  \tilde v_n(\vec p_1,\dots,\vec p_n;\vec x,\Lam) = v_n(\vec p_1,\dots,\vec p_n;\vec x) \\ + \sum_m\mathcal I(\Gam_n^{(m)};\vec p_1,\dots,\vec p_n;\vec x)
  \label{eq:graph_corr}
\end{multline}
for the vertex functions. We emphasize that the integrals $\mathcal I(\Gam_n)$ in general are functions of the outgoing wave vectors $\vec p_i$. To remain within the original model space (as defined by the function $F$), we have to reconstruct the functional form of the vertex $v_n$ neglecting terms involving higher orders of the wave vectors. The final step is to determine how the model parameters $x_i\to\tilde x_i$ change due to these graphical corrections by comparing the coefficients on both sides of Eq.~\eqref{eq:graph_corr}. While for simple vertex functions $v_n$ the $\tilde x_i$ can be read off directly, for functions that involve several outgoing wave vectors the problem can still be cast as a system of linear equations (Appendix~\ref{sec:renorm_parameters}).

\subsection{Wilson's shell renormalization}

The arguably most common scheme to implement the procedure sketched in Fig.~\ref{fig:renorm} is to consider an infinitesimal ``shell'' of wave vectors $k\in[\Lam/b,\Lam]$ through setting $b=1+\delta\ell$ and only consider contributions to linear order of $\delta\ell$. The first important consequence is that we only have to consider graphs with a single loop, and thus a single internal $\hat k$, since graphs with multiple loops are of higher order. Assuming that vertex functions are of the form $v_n=\sum_i x_iv_n^{(i)}$, from Eq.~\eqref{eq:graph_corr} we thus find
\begin{equation}
  \tilde x_i = (1+\psi_{x_i}\delta\ell)x_i
\end{equation}
with $\psi_{x_i}(\vec x)$ quantifying the graphical one-loop corrections for $x_i$ due to the non-linearities. The next step is to restore the cut-off $\Lam/b\to\Lam$ through rescaling, implying
\begin{multline}
  x_i' = b^{\Delta_{x_i}}\tilde x_i = (1+\delta\ell)^{\Delta_{x_i}}(1+\psi_{x_i}\delta\ell)x_i \\ \approx [1+(\Delta_{x_i}+\psi_{x_i})\delta\ell]x_i
  \label{eq:corr_param}
\end{multline}
to linear order. Wilson's flow equations thus read
\begin{equation}
  \partial_\ell x_i = (\Delta_{x_i}+\psi_{x_i})x_i = \beta_i(\vec x).
  \label{eq:wilson:flow}
\end{equation}

\subsection{Handling the integrals}
\label{sec:integrals}

At this point we have to face the integral $\int_{\hat k}$. The first step is to integrate over the internal frequency $\Om$, which only involves the bare propagators. The generalization of Eq.~\eqref{eq:S0} for the product $C_0(\Om,k)\prod_iG_0(s_i\Om,k_i)$ of the bare propagators inside the loop reads
\begin{multline}
  \int_{-\infty}^\infty\frac{\dd\Om}{2\pi} \frac{2Dk^\al}{\Om^2+[h(k)]^2}\prod_{i=1}^p\frac{1}{-s_i\im\Om+h(k_i)} = \\ \frac{Dk^\al}{h(k)}Q^{(p)}_{s_1\cdots s_n}\prod_{i=1}^p\frac{1}{h(k)+h(k_i)}
  \label{eq:I}
\end{multline}
with $\vec k_i$ the corresponding wave vector of the loop edge. Here, $s_i=\pm 1$ is the sign of $\hat k$ within $\hat k_i$ (remember that we set all external frequencies to zero). If all signs are equal then $Q^{(p)}=1$. For $p=2$ propagators and mixed signs one finds
\begin{equation}
  Q^{(2)}_{+-} = Q^{(2)}_{-+} = \frac{2h(k)+h(k_1)+h(k_2)}{h(k_1)+h(k_2)}
  \label{eq:Q}
\end{equation}
with similar but more complicated expressions for $p>2$ propagators.

The integral over the internal wave vector $\vec k$ is performed in spherical coordinates,
\begin{equation}
  \int\frac{\dd^d\vec k}{(2\pi)^d} = \frac{S_{d-1}}{(2\pi)^d}\IInt{k}{\Lam'}{\Lam}k^{d-1}\IInt{\theta}{0}{\pi}\sin^{d-2}\theta
  \label{eq:int:k}
\end{equation}
with polar angle $\theta$, $S_d\equiv2\pi^{d/2}/\Gam(d/2)$ the surface area of a unit hypersphere in $d$ dimensions, and $K_d\equiv S_d/(2\pi)^d$. Here, $\Gam(s)$ is the gamma function generalizing the factorial to non-integers. Useful angular integrals in the following are~\cite{gradshteyn2014table}
\begin{gather}
  \label{eq:S:0}
  S_{d-1}\IInt{\theta}{0}{\pi}\sin^{d-2}\theta = S_d, \\
  S_{d-1}\IInt{\theta}{0}{\pi}\sin^{d-2}\theta\cos\theta = 0, \\
  \label{eq:S:2}
  S_{d-1}\IInt{\theta}{0}{\pi}\sin^{d-2}\theta\cos^2\theta = \frac{S_d}{d}.
\end{gather}
Finally,
\begin{equation}
  \IInt{k}{\Lam/b}{\Lam}k^{d-1} h(k) = \Lam^d h(\Lam)\delta\ell + \mathcal O(\delta\ell^2)
  \label{eq:wilson:int}
\end{equation}
for any function $h(k)$ of the magnitude $k$.


\subsection{Illustrations}
\label{sec:illu}

\subsubsection{Liquid-gas phase separation and coexistence}
\label{sec:modelb}

\begin{figure*}
  \centering
  \includegraphics{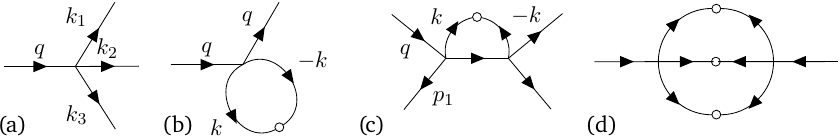}
  \caption{Relevant graphs of model B. (a)~Initial bare $3$-vertex. (b)~One-loop correction to the propagator. (c)~One-loop correction to the $3$-vertex involving two vertices. (d)~The first graphical correction to the noise strength is a two-loop integral.}
  \label{fig:modelB}
\end{figure*}

To demonstrate how dynamic renormalization works in practice, we first turn to the paradigmatic Model B describing the coexistence of two phases through the bulk free energy
\begin{equation}
  f(\phi) = \frac{a}{2}\phi^2 + \frac{u}{4}\phi^4.
  \label{eq:free_energy}
\end{equation}
For $a<0$ the free energy exhibits two minima at $\phi_\pm=\pm\sqrt{-a/u}$. Here, the order parameter field $\phi(\x,t)$ is related to the density (but could also be composition in case of a binary mixture). From now on we focus on the dynamics of a mass-conserving system (setting $\al=2$), i.e., any change of $\phi$ is due to a \emph{current} $\vec j(\x,t)$ with continuity equation
\begin{equation}
  \partial_t\phi + \nabla\cdot\vec j = \eta.
\end{equation}
Penalizing gradients (giving rise to an interfacial tension) then leads to the 
Ginzburg-Landau functional
\begin{equation}
  \mathcal F[\phi] = \Int{^d\x} \left[\frac{\kap}{2}|\nabla\phi|^2 + f(\phi)\right]
\end{equation}
and, assuming a current $\vec j=-\nabla\delta\mathcal F/\delta\phi$, to the evolution equation
\begin{equation}
  \partial_t\phi = \nabla^2\fd{\mathcal F}{\phi} = \nabla^2(a\phi+u\phi^3-\kap\nabla^2\phi) + \eta.
  \label{eq:phi:modelB}
\end{equation}
Let us first see when $u$ becomes irrelevant. Rescaling $a\phi+u\phi^3$ yields the scaling dimension $\Delta_u=\Delta_a-2\Delta_\phi$ and thus $\Delta^0_u=4-d$ with $\Delta^0_\phi=(d-2)/2$. In dimensions $d>4$ the non-linear term is irrelevant and the Gaussian fixed point is attractive. This changes for $d<4$ with $u(\ell)$ moving away from a small but non-zero initial $u_0$. We will now determine where it flows to. Model B has one non-zero vertex $v_3(q)=-uq^2$ [Fig.~\ref{fig:modelB}(a)], which implies that graphs can only be constructed from $3$-vertices.

\begin{figure}[t]
  \centering
  \includegraphics{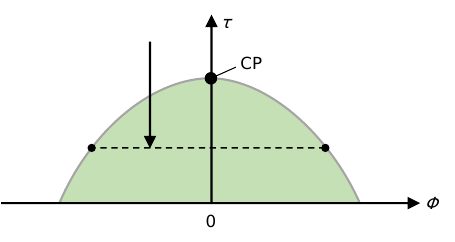}
  \caption{Sketch of the phase diagram of Model B following from the free energy density Eq.~\eqref{eq:free_energy}, where $\tau$ denotes the distance to the critical point (CP). The binodal (solid line) bounds the coexistence region. After a quench from the homogeneous into the coexistence region (arrow), the system becomes inhomogeneous with the coexisting values for the field $\phi$ given by the binodal (dashed line and symbols). The binodal ends in the critical point (CP), which is only reached for global $\phi=0$.}
  \label{fig:quench}
\end{figure}

We start with the graph in Fig.~\ref{fig:modelB}(b), which contributes
\begin{equation}
  \mathcal I(\Gam_1) = -3uq^2[G_0(\hat q)]^2\int_{\hat k}C_0(\hat k)
\end{equation}
to the propagator with multiplicity $|\Gam_1|=3$ since there are three ways to connect two out of three lines. The integral becomes
\begin{equation}
  \int_{\hat k}C_0(\hat k) = \int\frac{\dd^d\vec k}{(2\pi)^d}S_0(k) = K_d\Lam^dS_0(\Lam)\delta\ell
\end{equation}
with the static structure factor $S_0(k)$ given in Eq.~\eqref{eq:S0}. From $\tilde v_1(q)=v_1(q)+\mathcal I(\Gam_1)$ we find
\begin{multline}
  \tilde h(q) = h(q)\left[1 - 3uq^2\frac{K_d\Lam^dS_0(\Lam)}{h(q)}\delta\ell\right]^{-1} \\ \approx h(q) + 3uq^2 K_d\Lam^dS_0(\Lam)\delta\ell
\end{multline}
expanding again for small $\delta\ell\ll 1$. Plugging in $h(q)=q^2(a+\kap q^2)$, we can now read off the graphical corrections of the model parameters with intermediates $\tilde\kap=\kap$ (whence $\psi_\kap=0$) and
\begin{equation}
  \tilde a = a + 3u K_d\Lam^dS_0(\Lam)\delta\ell, \quad \psi_a = \frac{3uD}{a}\frac{K_d\Lam^d}{a+\kap\Lam^2}
\end{equation}
due to Fig.~\ref{fig:modelB}(b).

We now turn to the graph in Fig.~\ref{fig:modelB}(c). The multiplicity of this graph is: $|\Gam_3|=3$ (possibilities to insert the new vertex) $\times 2$ (remaining possibilities to insert a noise) $\times 3$ (possibilities to insert a noise in the new vertex). Reading from left to right vertex, we have
\begin{equation}
  \mathcal I(\Gam_3) = 18v_3(q)\int_{\hat k}C_0(\hat k)G_0(\hat q-\hat k-\hat p_1)v_3(\vec q-\vec k-\vec p_1)
  \label{eq:int:v3}
\end{equation}
leaving out the external lines (they are not part of the function $v_3$). We can immediately set $\hat p_1\to0$ inside the integral since also $\tilde v_3(q)=-\tilde uq^2$ will only depend on $q$. The frequency integral of $C_0(\hat k)G_0(\hat q-\hat k)$ then reads [cf. Eq.~\eqref{eq:I}]
\begin{multline}
  \int_{-\infty}^\infty\frac{\dd\Om}{2\pi}\frac{2Dk^2}{\Om^2+[h(k)]^2}\frac{1}{\im\Om+h(\vec q-\vec k)} \\ = \frac{Dk^2}{h(k)[h(k)+h(\vec q-\vec k)]}.
  \label{eq:int:CG}
\end{multline}
Since the pre-factor in Eq.~\eqref{eq:int:v3} is already $\propto q^2$ we can let $q\to0$ for the remaining terms inside the integral, which yields $\tilde v_3(q)=v_3(q)+\mathcal I(\Gam_3)$ and thus
\begin{equation}
  \tilde u = u - 9u^2D\frac{K_d\Lam^d}{(a+\kap\Lam^2)^2}\delta\ell, \quad \psi_u = -9uD\frac{K_d\Lam^d}{(a+\kap\Lam^2)^2}.
\end{equation}
A quick look at Fig.~\ref{fig:modelB}(d) reveals that the first correction to $D$ is already a two-loop integral and thus of order $(\delta\ell)^2$ with $\psi_D=0$.

\begin{figure}[b!]
  \centering
  \includegraphics{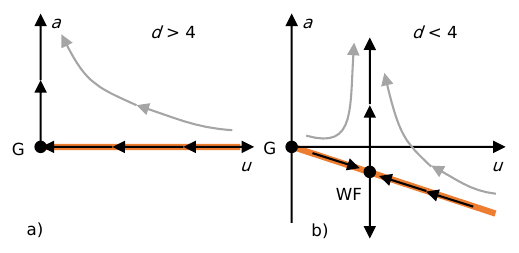}
  \caption{Sketch of the flow generated by Eq.~\eqref{eq:flow} for Model B in (a)~$d>4$ and (b)~$d<4$. Indicated are the Gaussian fixed point (G) and the Wilson-Fisher fixed point (WF). The highlighted orange line shows the critical manifold.}
  \label{fig:flow}
\end{figure}

For the final flow equations we introduce the reduced ``dimensionless'' model parameters
\begin{equation}
  \bar a \equiv \frac{a}{\kap\Lam^2}, \quad
  \bar u \equiv \frac{uD}{\kap^2}K_d\Lam^{d-4}
  \label{eq:aubar}
\end{equation}
leading to
\begin{equation}
  \partial_\ell\bar u = \left(\frac{\partial_\ell u}{u}+\frac{\partial_\ell D}{D}-2\frac{\partial_\ell\kap}{\kap}\right)\bar u = (\Delta_u+\Delta_D-2\Delta_\kap+\psi_u)\bar u
\end{equation}
through inserting their flow equations~\eqref{eq:wilson:flow}. Note how this choice removes the unknown exponents since with the scaling relations derived in Sec.~\ref{sec:scale} we find $\Delta_a-\Delta_\kap=2$ and $\Delta_u+\Delta_D-2\Delta_\kap=4-d=\eps$. Importantly, at this point the dimension $d$ has become simply a number and is not necessarily an integer. This freedom is exploited to introduce the small parameter $\eps$ with non-integer dimension $d=4-\eps$ close to the upper critical dimension at which the Gaussian fixed point becomes repulsive. The final flow equations then read
\begin{equation}
  \partial_\ell\bar a = 2\bar a + 3\bar u\frac{1}{\bar a+1}, \quad
  \partial_\ell\bar u = \left(\eps - 9\bar u\frac{1}{(\bar a+1)^2}\right)\bar u
  \label{eq:flow}
\end{equation}
expressed in the reduced parameters. Besides the Gaussian fixed point ($\bar a=\bar u=0$) these equations admit another fixed point of order $\eps$, the Wilson-Fisher (WF) fixed point located at $\bar u^\ast=\eps/9$ and $\bar a^\ast=-\eps/6$ to linear order~\cite{wilson72}. The resulting flow of the reduced parameters is plotted in Fig.~\ref{fig:flow}. For this model, the perturbative approach thus has been successful since the non-linear parameter $u\sim\eps$ remains small and does not run off.

To understand the flow around WF, we define $\tau\equiv\bar a-\bar a^\ast$ and $\delta\bar u=\bar u-\bar u^\ast$ measuring the distance to the critical point with new model parameters $\vec x=(\tau,\delta\bar u)^T$. The coupled linearized flow equations read
\begin{equation}
  \partial_\ell\left(\begin{array}{c}
    \tau \\ \delta\bar u    
  \end{array}\right) = \left(\begin{array}{cc}
    2-\tfrac{\eps}{3} & 3(1+\tfrac{\eps}{6}) \\ 0 & -\eps
  \end{array}\right)
  \left(\begin{array}{c}
    \tau \\ \delta\bar u    
  \end{array}\right)
\end{equation}
up to linear order of $\eps$. This matrix has two eigenvalues. The first is $\Delta_{\delta u}=-\eps$ and thus becomes irrelevant for $d<4$. As a consequence, the flow of $u$ is repelled from G and flows towards WF along its eigenvector. The second eigenvector is $\vec x_\tau=(1,0)^T$ so we identify its eigenvalue with $\Delta_\tau=2-\eps/3$. We expect the correlation length to diverge as $\xi\sim|\tau|^{-\nu}$ with exponent $\nu$. From the scaling dimensions (Sec.~\ref{sec:scale}) we immediately find $\Delta_\xi=-\nu\Delta_\tau=-1$ and thus $\nu=1/\Delta_\tau\approx\tfrac{1}{2}+\tfrac{\eps}{12}$. While strictly valid only close to $d=4$, the expansion in $\eps$ can be improved systematically and there is ample evidence that the WF controls the critical behavior down to $d=2$~\cite{leguillou85}. In the Supplemental Material, we provide a notebook to reproduce the results of this section~\cite{sm}.

\subsubsection{Kinetics of interfaces}
\label{sec:ckpz}

Whether interfaces are rough or smooth is of great technological importance, e.g., in the fabrication of semiconductors. The theoretical modeling of interfaces has a long tradition in statistical physics. Of particular importance is the minimal field theory by Kardar, Parisi, and Zhang (KPZ) for the large-scale dynamics of interfaces~\cite{kardar86}. KPZ defines a universality class beyond film growth and a number of exact results have been uncovered~\cite{prahofer04,takeuchi18}.

The field $\phi(\x,t)$ now describes an interface (or surface) in $d+1$ dimensions with ``substrate position'' $\x\in\mathbb R^d$. For a curl-free current we can write $\vec j=-\nabla\mu$ with an effective chemical potential $\mu$. Since $\mu$ is a scalar field, it can only involve rotationally invariant quantities. The lowest-order scalar that can be constructed for an interface is its local curvature $\nabla^2\phi$ with coefficient $\kap\geqslant0$ akin to a rigidity that resists bending the interface. Non-potential terms arise from an expansion of $\mu$ in even powers of $|\nabla\phi|$ (see Ref.~\cite{krug97} for a more comprehensive discussion) so that to lowest order one finds
\begin{equation}
  \mu_\text{cKPZ} = -\kap\nabla^2\phi + c_2|\nabla\phi|^2
  \label{eq:mu:cKPZ}
\end{equation}
with non-linear coefficient $c_2$ (conventionally $c_2=-\lam/2$ but we prefer $c_2$ for reasons that will become clear presently). Together with Eq.~\eqref{eq:phi}, this chemical potential defines the conserved KPZ (cKPZ) equation
\begin{equation}
  \partial_t\phi = \nabla^2(c_2|\nabla\phi|^2 - \kap\nabla^2\phi) + \eta
  \label{eq:phi:cKPZ}
\end{equation}
first studied by Sun, Guo and Grant~\cite{sun89}.

\begin{figure}[t]
  \centering
  \includegraphics{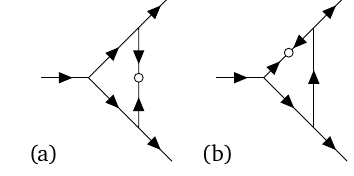}
  \caption{One-loop graphical corrections to the 2-vertex for the conserved KPZ [Eq.~\eqref{eq:phi:cKPZ}].}
  \label{fig:graphs_ckpz}
\end{figure}

Invariance of Eq.~\eqref{eq:phi:cKPZ} under scaling yields the relations
\begin{equation}
  \Delta_\kap = z-4, \quad \Delta_2 = z-4-\Delta_\phi, \quad \Delta_D = 2\Delta_\phi - d + z
\end{equation}
between the scaling dimensions. Unlike KPZ, Eq.~\eqref{eq:phi:cKPZ} does not exhibit Galilean invariance and, therefore, the non-linear coefficient $c_2$ receives graphical corrections from the graphs shown in Fig.~\ref{fig:graphs_ckpz} although their contributions cancel each other at one-loop~\cite{janssen97}.

Let us calculate the remaining corrections $\psi_i$. The only non-zero vertex is $v_2(\vec k_1,\vec k_2)=-c_2q^2\vec k_1\cdot\vec k_2$. For the graph shown in Fig.~\ref{fig:graph}(c), we have already calculated the frequency integral in Eq.~\eqref{eq:int:CG}. Plugging this result with $h(k)=\kap k^4$ into Eq.~\eqref{eq:int:v2}, we find
\begin{equation}
  \begin{aligned}
  \mathcal I(\Gam_1) = & -\frac{4c_2^2D q^2}{\kappa^2} [G_0(\hat q)]^2\\
  & \int\frac{\dd^d\vec k}{(2\pi)^d} \frac{\vec k\cdot(\vec q-\vec k)(\vec q\cdot\vec k)|\vec k - \vec q|^2}{k^2(k^4+|\vec q-\vec k^4|)}.
  \label{eq:ckpz:int}
  \end{aligned}
\end{equation}
Performing the angular integrals [Eqs.~\eqref{eq:S:0} and~\eqref{eq:S:2}] and reading off $\tilde\kap$, we find
\begin{equation}
  \psi_\kap = \frac{2c_2^2D}{d\kap^3}K_d\Lam^{d-2} = \frac{2\bar c_2^2}{d}.
  \label{eq:kpz:kap}
\end{equation}
Importantly, since the lowest order is $q^4$ this graph only corrects $\kap$ but does not generate a correction to $a$, which remains zero. Here we have defined the sole dimensionless parameter $\bar c_2^2\equiv c_2^2D\kap^{-3}K_d\Lam^{d-2}$. We also see that the corresponding graphical correction for the noise term [Fig.~\ref{fig:graph}(e)] is $\mathcal{O}(q^4)$. The original noise correlations [Eq.~\eqref{eq:K}] are proportional to $q^2$ whence $\psi_D=0$.

The flow equation for the single dimensionless model parameter now reads
\begin{equation}
  \begin{aligned}
  \partial_\ell\bar c_2 & = \left(\frac{\partial_\ell c_2}{c_2}+\frac{1}{2}\frac{\partial_\ell D}{D}-\frac{3}{2}\frac{\partial_\ell\kap}{\kap}\right)\bar c_2  \\
  & = \left(\Delta_2 + \frac{1}{2}\Delta_D - \frac{3}{2}\Delta_\kap- \frac{3}{2}\psi_\kap\right)\bar c_2.
  \end{aligned}
\end{equation}
Plugging in the relations for the scaling dimensions and the graphical correction [Eq.~\eqref{eq:kpz:kap}], we obtain the closed flow equation
\begin{equation}
  \partial_\ell\bar c_2 = \left(\frac{2-d}{2}-\frac{3\bar c_2^2}{d}\right)\bar c_2.
\end{equation}
We recover the Gaussian fixed point at $\bar c_2=0$, which is attractive for $d>2$ and repulsive for $d<2$. In $d<2$ dimensions there is a second perturbative fixed point at $|\bar c_2^\ast|=\sqrt{\frac{\eps}{3}}$ to lowest order in $\eps=2-d$, which is connected to a dynamic phase transition (``roughening transition''). For a discussion of its physical effects on the growth of interfaces, we refer to the literature, e.g. Ref.~\cite{sun89}. Again, we provide a notebook to reproduce the results of this section~\cite{sm}.


\section{Active Model B+}
\label{sec:ambp}

\begin{figure*}
  \centering
  \includegraphics{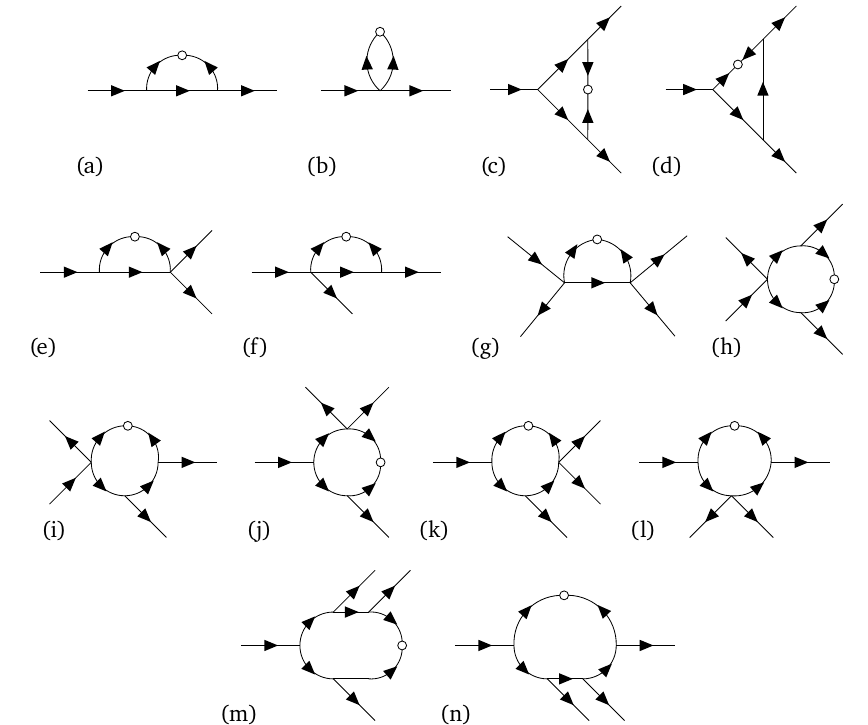}
  \caption{All possible one-loop graphs for a model with 2-vertices and 3-vertices. These contribute graphical corrections to (a,b)~the propagator, (c-f)~2-vertices, and (g-n) 3-vertices.}
  \label{fig:graphs}
\end{figure*}

\subsection{Definition of the model}

Although at first glance the physics of Model B and cKPZ are quite different, mathematically both fall into the same class of scalar theories. The most general form including all possible terms at order $\phi^2$ and $\nabla^4$ reads
\begin{multline}
  \partial_t \phi = \nabla^2 (a \phi+b \phi^2+u\phi^3-\kappa \nabla^2 \phi) + \\
  \frac{c_1}{2}\nabla^4 \phi^2+c_2\nabla^2 |\nabla \phi|^2 - c_3 \nabla [ (\nabla^2\phi)\nabla \phi ] + \eta
  \label{eq:phi:ambp}
\end{multline}
with three new model parameters $c_i$. Note that at next order this model only includes $u\nabla^2\phi^3$ although in principle more terms at that order are admissible. Equation~\eqref{eq:phi:ambp} has been coined Active Model B+ (AMB+)~\cite{tjhung18} and, so far, it has been discussed mostly in the context of active Brownian particles. Clearly, Model B follows setting $c_1=c_2=c_3=0$ and cKPZ follows setting $a=b=u=c_1=c_3=0$.

The central lesson from the $\eps$-expansion of Model B in Sec.~\ref{sec:modelb} has been that perturbative corrections give rise to a new critical fixed point, the Wilson-Fisher (WF) fixed point, which detaches from the Gaussian fixed point and controls the critical behavior for $d<4$. The central question that we address in the following is whether such a similar scenario occurs in $d=2+\eps$ dimensions when including the additional derivatives of Active Model B+.

In contrast to the illustrations of Sec.~\ref{sec:illu}, the presence of both 2-vertices and 3-vertices now implies a large number of graphs to consider, which are shown in Fig.~\ref{fig:graphs}. While $v_3(q)=-uq^2$, the vertex function for 2-vertices
\begin{multline}
  v_2(\vec p_1,\vec p_2| \vec q) = -b\vec q^2 + \frac{c_1}{2}\vec q^4 + c_2 \vec p_1 \cdot \vec p_2 q^2 \\ -\frac{c_3}{2}(\vec p_2^2 \vec q\cdot \vec p_1+\vec p_1^2 \vec q\cdot \vec p_2)
  \label{eq:v2}
\end{multline}
has become considerably more complex compared to our illustrative cases and requires to handle more involved integrals\footnote{We have developed a python package that handles graph construction and mapping to integrals symbolically, which can be found at \url{https://github.com/us-itp4/restflow}. See also the Jupyter notebooks provided in the Supplemental Material~\cite{sm}.}.

We point out that there is one combination $c_3=0$ and $c_1+2c_2=0$ of the new model parameters for which we can restore an effective free energy
\begin{equation}
  \mathcal F[\phi] = \Int{^d\x} \left[\frac{\kap-c_1\phi}{2}|\nabla\phi|^2 + f(\phi)\right].
\end{equation}
In the following, we will denote this condition as the ``equilibrium line'' in parameter space.

\subsection{The cubic coefficient $b$ flows}
\label{sec:cubic}

Before we embark on deriving the flow equations for AMB+, we note that, unlike for Model B, even if we start with $b=0$ there are now graphical corrections that cause $b$ to flow away from zero. Specifically, the one-loop graphical correction becomes (see appendix~\ref{sec:graphs_ambp} for the detailed calculation)
\begin{multline}
  \psi_b b = \frac{3D\kap u}{2d b\kap^2}K_d\Lambda^{d-2} [2c_1d-2c_2d-c_3(d-2)] \\ +\frac{Dc_1^2}{d b\kap^3}K_d \Lam^d[-2c_3+d(2c_2+c_3)],
   \label{eq:psi:b}
\end{multline}
which is non-zero if $u\neq 0$ and any $c_i\neq 0$, or if $u=0$ and $c_1\neq0$ and either $c_2\neq0$ or $c_3\neq0$. For non-zero $b$, the 2-vertex function Eq.~\eqref{eq:v2} thus contributes to further $b$-dependent terms in the graphical corrections of the other (dimensionless) model parameters due to graphs including 2-vertices.

To elucidate the role of the cubic coefficient $b$, it is instructive to briefly return to Model B. The modified bulk free energy 
\begin{equation}
  f(\phi) = \frac{a}{2}\phi^2 + \frac{b}{3}\phi^3 + \frac{u}{4}\phi^4
  \label{eq:energy_cubic}
\end{equation}
including a cubic term $\phi^3$ can be interpreted as an off-critical quench. A shift $\phi\to\phi+\phi_0$ of the field allows to eliminate the cubic coefficient $b\to \hat{b}=0$ through setting $\phi_0=-b/(3u)$. This shift modifies $a\to \hat{a}=a-b^2/(3u)$ and also introduces a linear term $\propto\phi$ corresponding to an external field that explicitly breaks the symmetry $\phi\to-\phi$ of Eq.~\eqref{eq:free_energy}. Through this shift, we thus transform the phase diagram to the one sketched in Fig.~\ref{fig:quench} and can now reach the critical fixed point through tuning $a$ along the line $\phi=0$. This interpretation hinges on the physical nature of the order parameter. Note that the cubic non-linearity in the free energy is not always eliminated through such a shift. For example, it is necessary for modeling liquid crystals~\cite{gennes93} and can result in interesting behaviors, such as a weak first order transition when fluctuations are taken into account~\cite{mukherjee98a}.

On the other hand, without the shift keeping $b\neq0$, one finds that the flow of Model B diverges. Based on Fig.~\ref{fig:quench}, we interpret the flow to approach the trivial fixed points with $a\rightarrow \pm \infty$ (with $u$, $b$ roughly constant) corresponding to the ``pure'' phases (e.g. gas or liquid). Therefore, the only other non-trivial fixed point besides G is WF at $b=0$ and fine-tuning $a$. Following this argument thus fixes the zero value of the order parameter field $\phi$ to the critical manifold. For more details of Model B with $b\neq0$, see Appendix~\ref{sec:modelb_un}.

\subsection{Generation of relevant higher-order terms}
\label{sec:higher}

\begin{figure}[t]
  \centering
  \includegraphics{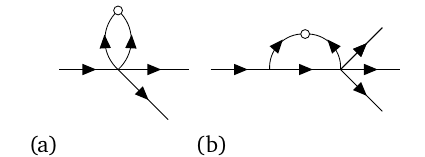}
  \caption{Examples of one-loop graphs with a 4-vertex that contribute graphical corrections to (a) the 2-vertex function and (b) the 3-vertex function.}
  \label{fig:graphs_4_vert}
\end{figure}

Including 2-vertices (absent in standard Model B) allows to construct one-loop graphs that correspond to higher-order non-linear terms of the form $\nabla^2\phi^n$ with $n\geq3$ that are not part of Eq.~\eqref{eq:phi:ambp}. This is reminiscent of the field-theoretic renormalization of the annihilation-fission process~\cite{howard97}. The graphical correction of the corresponding coefficient is proportional to $c_1^n f(c_2, c_3)$, where $f(c_2, c_3)$ is a linear homogeneous function of $c_2$ and $c_3$ [cf. Eq.~\eqref{eq:psi:b} for $u=0$]. Therefore, these terms are generated only for $c_1\neq0$ and either $c_2\neq0$ or $c_3\neq0$. Since they correspond to new $n$-vertices, new one-loop graphs can be constructed (see Fig.~\ref{fig:graphs_4_vert} for two examples) that contribute to further corrections to the model parameters. Importantly, the flow of the corresponding model parameters couples with the rest of the flow.

``Naive'' dimensional analysis with $\Delta_\phi^0=(d-2)/2$ predicts that higher-order terms $\nabla^2\phi^n$ become irrelevant for $d>2\frac{n+1}{n-1}$ with $n\geq4$. Therefore, we expect terms with $n>5$ to be irrelevant for $d\geq3$. However, as explained in Section~\ref{sec:scale}, such a dimensional analysis neglects the anomalous dimensions $\eta$ and, in principle, we might need to consider these terms in the one-loop analysis.

\subsection{Renormalization flow close to the WF}
\label{sec:ambp_flow}

\subsubsection{Full flow equations}
\label{sec:noshift}

With these considerations in mind, we now turn to the flow equations for the full AMB+ [Eq.~\eqref{eq:phi:ambp}]. Similarly to Sec.~\ref{sec:modelb}, we employ the reduced dimensionless model parameters $\bar a$ and $\bar u$ defined in Eq.~\eqref{eq:aubar} and also define
\begin{gather}
  \label{eq:bbar}
  \bar b \equiv \frac{b D^{1/2}}{\kappa^{3/2}} K_d^{1/2} \Lambda^{d/2-3}, \\
  \label{eq:cbar}
  \bar c_i \equiv \frac{c_i D^{1/2}}{\kap^{3/2}}K_d^{1/2}\Lam^{d/2-1}.
\end{gather}
The final flow equations read
\begin{gather}
  \label{eq:flow:a_amb}
  \partial_\ell \bar a = \left(2+\psi_a - \psi_\kappa\right) \bar a, \\
  \label{eq:flow:u_amb}
  \partial_\ell \bar u = (4-\eps+\psi_u-2\psi_\kappa) \bar{u}, \\
  \label{eq:flow:b_amb}
  \partial_\ell \bar b = \left(\frac{6-d}{2}-\frac{3}{2}\psi_\kappa+\psi_b \right) \bar{b}, \\
  \label{eq:flow:c_amb}
  \partial_\ell \bar c_i = \left(-\frac{d-2}{2}-\frac{3}{2}\psi_\kappa+\psi_i \right) \bar c_i.
\end{gather}
The graphical corrections $\psi_x$ up to lowest order of the model parameters and up to $\mathcal{O}(\eps)$ are obtained in Appendix~\ref{sec:graphs_ambp}. Naive dimensional analysis would suggest that the $c_i$ coefficients become irrelevant above two dimensions. To proceed, we expand these evolution equations around $d=2+\eps$ with $\eps$ assumed to be small.

\begin{table}[b!]
  \begin{tabular}{c|c|c|c|c}
  \hline
             & WF                             & F$_\pm$                          & K$^1_\pm$            & K$^2_\pm$            \\ \hline\hline
  $\bar a$   & $-\frac{1}{4}+\frac{31}{48}\eps$ & $-\frac{1}{4}+\frac{31}{48}\eps$ & $-1.9+0.2\eps$      & $1.2+0.45\eps$     \\[0.08cm] \hline
  $\bar u$   & $\frac{2}{9}-\frac{1}{9}\eps$  & $\frac{2}{9}-\frac{1}{27}\eps$   & $-0.25-0.1\eps$      & $0.07+0.03\eps$      \\[0.08cm] \hline
  $\bar b$   & $0$                            & $\pm \eps^{1/2}$                 & $\pm1.2\pm0.12\eps$ & $\pm0.7\pm0.24\eps$ \\[0.08cm] \hline
  $\bar c_1$ & $0$                            & $\mp \frac{1}{3}\eps^{1/2}$      & $\pm1.2\pm0.12\eps$ & $\mp0.66\mp0.32\eps$ \\[0.08cm] \hline
  $\bar c_2$ & $0$                            & $\mp \frac{1}{3}\eps^{1/2}$      & $\mp0.6\mp0.06\eps$ & $\pm0.33\pm0.16\eps$ \\[0.08cm] \hline
  $\bar c_3$ & $0$                            & $\pm 2\eps^{1/2}$                & $0$                  & $0$                  \\[0.08cm] \hline
  \end{tabular}
  \caption{Fixed points of the truncated flow equations for AMB+ with $\eps=d-2$.}
  \label{table:fixed_pts_ambp}
\end{table}

In Table~\ref{table:fixed_pts_ambp}, we show the fixed points admitted by the flow equations~\cite{sm}. One fixed point agrees with the extrapolation of the result of Sec.~\ref{sec:modelb}, and we identify this fixed point with the WF although the values for $\bar u$ and $\bar a$ are now of order $\mathcal O(1)$. However, for small values of the other parameters their flow decouples from $\bar a$ and $\bar u$. Based on the proposition that the WF critical point persists to lower dimensions, we thus probe the influence of the additional model parameters and whether additional fixed points influence the critical behavior.

There is one pair F$_\pm$ of perturbative fixed points in the sense that they merge with WF in the limit $\eps\to0$. They are only accessible for $d\geqslant2$. However, a stability analysis shows that these fixed points exhibit complex eigenvalues. The corresponding eigenvectors overlap with all axes, preventing the flow to converge towards F$_\pm$ along their irrelevant directions, making them effectively unreachable by the flow. Note that this situation is reminiscent of Model B with $b\neq0$, where we also find a pair of fixed points with complex eigenvalues (Appendix~\ref{sec:modelb_un}). In this case, the new pair of fixed points is also not reachable and it just modifies the boundary between $\bar a \rightarrow \infty$ and $\bar a \rightarrow -\infty$. Indeed, it appears that F$_\pm$ is likely its counterpart for AMB+ since they both exhibit spiral behavior along $\bar a$ and $\bar u$ and converge towards WF for $d\rightarrow2$. There are two more non-perturbative pairs of fixed points emerging from the truncated flow equations, which do not alter the flow close enough to the WF due to their large $\bar c$ values.

Linear stability analysis of the WF reveals that in $d\geqslant 2$ all of its eigenvalues are non-positive apart from one, the corresponding eigenvector of which points along the $\bar a$ axis. To obtain further insights, we numerically integrate the flow equations in $d>2$ for non-zero $\bar b$ and $\bar c_n$, and we observe for $\bar a$ fine-tuned to its critical value that the flow converges towards the WF fixed point. Interestingly, the WF now exhibits a conjugate pair of complex eigenvalues with negative real part in $d>2$, which leads to the flow spiraling down into the WF on the hyperplane spanned by $(\bar b, \bar c_1, \bar c_2)$, and thus does not affect the large-scale physics. For $d=2$, there is a Hopf bifurcation and the system oscillates on that hyperplane around the WF.

\subsubsection{Constraining $b$ to zero}
\label{sec:shift}

In the previous section (Sec.~\ref{sec:noshift}), we have seen that the flow of model parameters leads to inconsistencies reminiscent of conventional Model B with a non-zero cubic coefficient $b$. While for Model B we can nevertheless access the critical manifold through shifting the origin of the field, now Eq.~\eqref{eq:psi:b} implies 
\begin{equation}
  \tilde b = \left.\psi_b b \right|_{b=0} \delta\ell \neq 0
\end{equation}
even for $b=0$ due to non-zero graphical corrections. Moreover, a shift $\phi\to\phi+\phi_0$ of the field also affects $\kap\to\hat{\kap}=\kap-c_1\phi_0$. This effect has been overlooked by previous studies~\citep{caballero18a,speck22} and we explicitly describe the field-shift procedure here.

Shifting the field after adding the graphical corrections leads to new intermediate model parameters $\hat{\vec x}$ with
\begin{equation}
  \hat\kap = \tilde\kap + \frac{\tilde c_1\tilde b}{3\tilde u}, \quad
  \hat a = \tilde a - \frac{\tilde b^2}{3\tilde u}, \quad \hat b = 0,
\end{equation}
while the other model parameters are not affected by the shift. The correction to $a$ is of order $\delta\ell^2$ and thus can be dropped, $\hat a=\tilde a$. For $\kap$, we absorb the additional term up to order $\mathcal{O}(\delta\ell)$ into
\begin{equation}
  \hat\psi_\kap \equiv \left(\psi_\kap + \frac{c_1}{3u\kap}\psi_b b\right)_{b=0}
\end{equation}
with all other graphical corrections $\psi_x$ unchanged. The remaining flow equations are the same as in \cref{eq:flow:a_amb,eq:flow:u_amb,eq:flow:c_amb} with $\psi_\kap$ replaced by $\hat\psi_\kap$. We again expand around $d=2$ with $\eps=d-2$.

Since we have to apply the field shift in every infinitesimal step to constrain $b$, the value of $\phi_0$ keeps changing and flows as $\ell$ increases. Summing these changes implies
\begin{equation}
  \partial_\ell\phi_0 = \Delta_\phi\phi_0 - \frac{1}{3u}\left.\psi_b b\right|_{b=0}
\end{equation}
after rescaling, for which we employ the scaling dimension $\Delta_\phi$ of the field. To eliminate trivial effects we introduce the scaled field origin
\begin{equation}
  \bar\phi_0 \equiv \frac{\phi_0\kap^{1/2}}{D^{1/2}}K_d^{-1/2}\Lam^{1-d/2},
\end{equation}
for which we find
\begin{equation}
  \partial_\ell\bar\phi_0 = \left(\frac{d-2}{2}+\frac{1}{2}\hat\psi_\kap\right)\bar\phi_0 + \Psi
\end{equation}
with $\Psi(\bar{\vec x})=-\bar c_1+\bar c_2+\frac{\eps}{4}\bar c_3$ to lowest order. Since $\bar\phi_0$ does not influence the flow of the other model parameters, we can go to a fixed point and inquire about the fate of $\bar\phi_0$. The differential equation is $\partial_\ell\bar\phi_0=\al\bar\phi_0+\Psi$ with $\al\equiv(\eps+\hat\psi_\kap)/2$ evaluated at the fixed point. This equation is solved by ($\al\neq 0$)
\begin{equation}
  \bar\phi_0(\ell) = \frac{\Psi}{\al}\left(e^{\al\ell}-1\right)
\end{equation}
and $\bar\phi_0(\ell)=\Psi\ell$ for $\al=0$, setting the initial value $\bar\phi_0(\ell=0)=0$ to zero. There are two possible outcomes in the limit $\ell\to\infty$: (i) $\bar\phi_0$ diverges for $\al>0$ and $\Psi\neq 0$, or (ii) it goes to zero for $\Psi=0$ or $\al<0$ and non-zero $\Psi$. Clearly, for G and WF we have $\Psi=0$ in any dimension and thus $\bar\phi_0=0$ at all scales.

\begin{table}[t]
  \begin{tabular}{c|c|c|c}
  \hline
             & WF                             & F$_\text{eq}$                   & K$_\text{eq}$                                                                                    \\ \hline\hline
  $\bar a$   & $-\frac{1}{4}+\frac{3}{32}\eps$ & $- \frac{1}{4}-\frac{\eps}{48}$       & $-\frac{3}{10}+\frac{7\eps}{25}$                                            \\[0.1cm] \hline
  $\bar u$   & $\frac{2}{9}-\frac{1}{9}\eps$  & $\frac{2}{9} - \frac{\eps}{27}$ & $\frac{2}{9}+\frac{5 \eps}{81}$                                   \\[0.1cm] \hline
  $\bar c_1$ & $0$                            & $\pm \frac{2 \sqrt{-\eps}}{3}$  & $\pm \frac{2 \sqrt{21} \sqrt{-\eps}}{9}$  \\[0.1cm] \hline
  $\bar c_2$ & $0$                            & $\mp\frac{\sqrt{-\eps}}{3}$     & $\mp \frac{\sqrt{21} \sqrt{-\eps}}{9}$                              \\[0.1cm] \hline
  $\bar c_3$ & $0$                            & $\mathcal{O}(\eps^2)$           & $0$                                                                                             \\[0.1cm] \hline
  \end{tabular}
  \caption{Fixed points of AMB+ to lowest order when constraining the flow to $b=0$. The two fixed points F$_\text{eq}$ and K$_\text{eq}$ merge with WF as $\eps\to0$ and lie on the equilibrium line $\bar c_1+2\bar c_2=0$.}
  \label{table:fixed_pts_ambp_shift}
\end{table}

We now determine the fixed points of the constrained flow, which are summarized in Table~\ref{table:fixed_pts_ambp_shift}~\cite{sm}. We again find a fixed point that we identify with WF. In addition, there are two fixed points F$_\text{eq}$ and K$_\text{eq}$ on the equilibrium line $\bar c_1+2\bar c_2=0$ that require $\eps\leqslant0$ ($d\leqslant2$) and that seem to merge with WF as $\eps\to0$. However, considering the value $\bar\phi_0$ of the field origin, we note that $\al\propto\eps$ and thus even though $\al<0$, its value $\Psi/\eps$ diverges as $\eps\to0$ and these fixed points are inaccessible. For $d\geqslant2$ there are no further fixed points with $\bar u\neq0$ apart from the WF. By numerically integrating the flow equations, we observe that the flow moves in general towards the WF. For $d=2$, there is a marginal direction and the system converges towards the WF logarithmically slow.

\subsubsection{Summary}

The conclusion from our one-loop analysis is that for $u>0$ the Wilson-Fisher (WF) fixed point is the only critical fixed point for AMB+ close to two dimensions, and that it determines its large-scale physics. While we found other fixed points in the vicinity of the (one-loop) WF, we conclude that they are artifacts and do not correspond to physically reachable critical points. Moreover, we find that the model parameters $\bar c_i$ are irrelevant at large scales since the flow converges towards the WF. This conclusion holds for not too large model parameters. Going further away from the WF there might be other relevant fixed points but we have argued that the presence of 2-vertices in AMB+ lead to the generation of new, possibly relevant, order parameters and, therefore, AMB+ is not ``closed'' at the level of one-loop dynamic renormalization. Notebooks with the detailed calculations of the flow equations can be found in the Supplemental Material~\cite{sm}.


\section{Perturbatively accessible field theories}
\label{sec:models}

As outlined in Sec.~\ref{sec:higher}, for $c_1\neq 0$ and either $c_2\neq0$ or $c_3\neq0$, additional terms to Eq.~\eqref{eq:phi:ambp} might be generated. In this section, we consider two models that fall into the AMB+ class of active field theories but that do not fulfill this condition and thus are ``closed'' in the sense that no additional non-linear terms are generated during the renormalization procedure.


\subsection{Kinetics of interfaces revisited}
\label{sec:mod_ckpz}

We first return to the conserved KPZ equation discussed in Sec.~\ref{sec:ckpz}. Recently, Caballero and coworkers have argued that Eq.~\eqref{eq:mu:cKPZ} is not the most general form and that geometric arguments admit another contribution to the surface current
\begin{equation}
  \vec j = -\nabla\mu_\text{cKPZ} + c_3(\nabla^2\phi)\nabla\phi
  \label{eq:j:cKPZ+}
\end{equation}
that describes ``blind geodesic jumps''~\cite{caballero18a}. Consequently, this extension of cKPZ called cKPZ+ involves a current with non-vanishing curl.

In fact, general arguments~\cite{krug97} indicate that, once detailed balance is broken on the level of microscopic particle jumps, an expansion of the surface current should also admit odd powers of $\nabla\phi$. Already to linear order this introduces an ``Edwards-Wilkinson'' diffusion term $\vec j_\text{EW}=-a\nabla\phi$ with coefficient $a$ that can take both signs. In addition, we assume that the interface growth is confined by an external potential (e.g., through the presence of a substrate) that breaks translational invariance and implies a field-dependent rigidity $\kap(\phi)$, which to lowest order of $\phi$ is expanded as $\kap(\phi)=\kap-c\phi$. Moreover, we focus on the special case where the coefficient of the square-gradient term also equals $c$ and we absorb both terms into
\begin{equation}
  c|\nabla\phi|^2 + c\phi\nabla^2\phi = \frac{c}{2}\nabla^2\phi^2.
\end{equation}
The final evolution equation becomes
\begin{equation}
  \partial_t\phi = \nabla^2\left(a\phi + \frac{c}{2}\nabla^2\phi^2 - \kap\nabla^2\phi\right) + \eta
  \label{eq:modckpz}
\end{equation}
and thus corresponds to AMB+ [Eq.~\eqref{eq:phi:ambp}] with $c_1=c$ and $b=u=0$ as well as $c_2=c_3=0$. As argued in Sec.~\ref{sec:higher}, the later condition is required to prevent corrections to $u$ and $b$ in the presence of a non-vanishing coefficient $c_1\neq 0$.

The flow equations for the dimensionless parameters $\bar a$ [Eq.~\eqref{eq:aubar}] and $\bar c$ [Eq.~\eqref{eq:cbar}] read
\begin{gather}
  \label{eq:flow:mod_ckpz_a}
  \partial_\ell \bar a = \left(2+\psi_a -\psi_\kappa\right) \bar a, \\
  \label{eq:flow:mod_ckpz_c}
  \partial_\ell \bar c = \left(-\frac{d-2}{2}-\frac{3}{2}\psi_\kappa+\psi_c \right) \bar c
\end{gather}
with graphical corrections
\begin{equation}
  \psi_a = 0, \quad
  \psi_\kap = -\frac{\bar c ^2}{2(\bar a + 1)^2}, \quad
  \psi_c = \frac{\bar c ^2}{(\bar a + 1)^3}
\end{equation}
calculated from the one-loop graphs in Fig.~\ref{fig:graphs}(a,c,d)~\cite{sm}. Visual inspection of Fig.~\ref{fig:graph}(e) reveals that the corresponding graphical correction for the noise term is of $\mathcal{O}(q^8)$ and thus $\psi_D=0$.

\begin{figure}[t]
  \centering
  \includegraphics{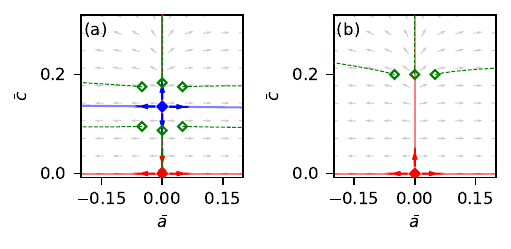}
  \caption{Modified cKPZ. Streamline plot of the flow equations \eqref{eq:flow:mod_ckpz_a} and \eqref{eq:flow:mod_ckpz_c} for the one-loop graphs in Fig.~\ref{fig:graphs}(a, c, d): (a)~$\eps=0.1$ ($d>2$) and (b)~$\eps=0$ ($d=2$). For $\eps<0$ ($d<2$), the flow is qualitatively similar to $\eps=0$ shown in (b). Note that we plot only for $\bar c>0$ since the flow equations are invariant under the transformation $\bar c \rightarrow -\bar c$. The closed red and blue dot corresponds to the Gaussian and Wilson-Fisher fixed point, respectively, and the green open symbols indicate the initial points of a few trajectories obtained through integrating the flow equations (dotted lines). The eigenvectors of the fixed points are the straight lines denoted by the same color as the fixed point.}
  \label{fig:mod_ckpz}
\end{figure}

\begin{figure}[t]
  \centering
  \includegraphics{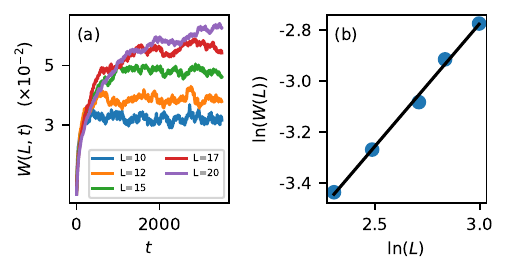}
  \caption{Growth of the width of the interface $W(L,t)$ versus time for the modified cKPZ [Eq.~\eqref{eq:modckpz}] in $d=3$, for $c=0.3$, $a=0$ for different system sizes. The lines and the dots are the averages over 200 trajectories. (b) The saturated width (the width at the last timestep from (a)) for different system sizes. The black line is obtained from linear regression and has slope $\lambda=0.965\pm0.037$ (c.f. for the Gaussian fixed point $\lambda=2\Delta_\phi=1$).}
  \label{fig:ckpz_3d}
\end{figure}

We plot the flow in the plane $(\bar a, \bar c)$ for $d>2$ [Fig.~\ref{fig:mod_ckpz}(a)] and $d=2$ [Fig.~\ref{fig:mod_ckpz}(b)]. The Gaussian fixed point G at $(0,0)$ is repulsive along the $\bar a$ direction for any $d$ and attractive (repulsive) for $d>2$ ($d\leqslant2$) along the $\bar c$ direction. In $d>2$, we also find a pair of perturbative fixed points A$_\pm$ at $(0,\pm\sqrt{2\eps/7})$, which are repulsive. We numerically integrate the flow and we find that for initially non-zero values of $\bar a$, it runs off to infinity in any dimension. Importantly, $\bar c$ also runs off to infinity except in $d>2$ with initial $|\bar c|\leqslant \sqrt{2\eps/7}$, for which it converges to $0$.

To corroborate the analytical results, we perform numerical finite-difference simulations of Eq.~\eqref{eq:modckpz} employing the Python package py-pde~\cite{zwicker20,sm}. We track the time evolution of the width $W$ of the interface given by $W(L,t)=\mean{\phi(\mathbf{x},t)^2}$, where $\mean{\cdot}$ denotes the average over noise histories and $\mean{\phi}=0$. From the scaling of $W$, the exponent $\Delta_\phi$ (see Sec.~\ref{sec:ckpz}) is calculated based on $W(L)\propto L^{2\Delta_\phi}$, where $W(L)$ is the saturated width, i.e., $W(L,t)$ for $t\rightarrow\infty$ (for more details on the procedure, see Ref.~\cite{halpin-healy95}).

To access the Gaussian fixed point, we perform numerical simulations for a three-dimensional substrate ($d=3$). In Fig.~\ref{fig:ckpz_3d}, we plot the numerical results for a parameter region in which the renormalization flow is supposed to approach the Gaussian fixed point. In this case, we find $\Delta_\phi=0.483\pm0.019$, which is in good agreement with $\Delta_\phi=(d-2)/2$ (see Sec.~\ref{sec:scale}). For $d=3$ and large enough values of $c$, the simulations develop a numerical instability (the profile becomes very steep within a few time steps). For $d=1,2$ with $c\neq0$, this is always the case. We interpret this instability to indicate non-trivial behavior and the potential crossover to a strong-coupling regime. It has indeed been observed in the strong-coupling regime of the KPZ equation~\cite{moser91} and necessitates different numerical approaches such as performing a Cole-Hopf transformation~\cite{beccaria94} or a special fitting ansatz~\cite{ala-nissila94}. Note that without changing the final conclusions, here we set $a=0$ to accelerate the saturation of the width or the explosion of the trajectories.


\subsection{Neural network model}
\label{sec:neural} 

\begin{figure*}[t]
  \centering
  \includegraphics{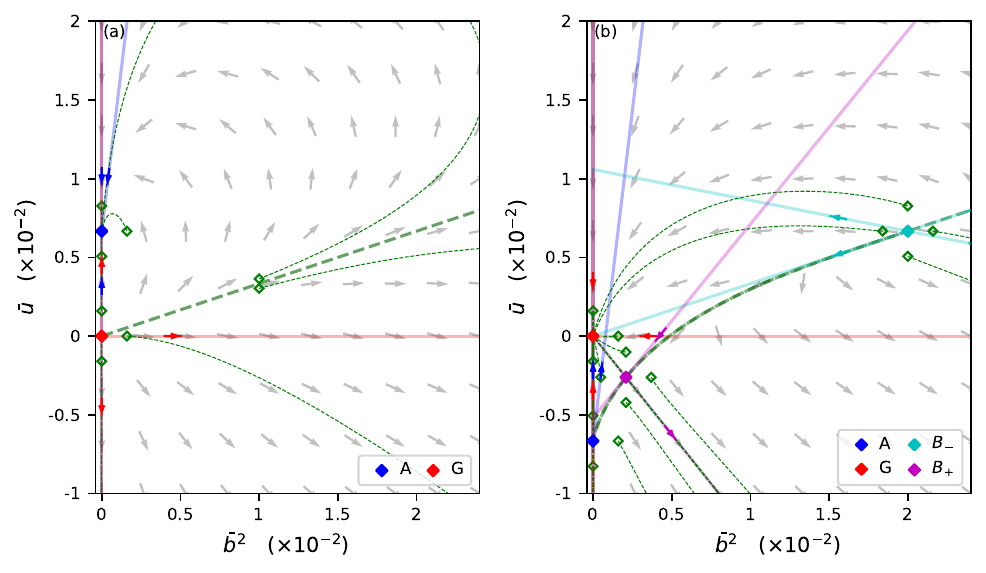}
  \caption{Neural network. Streamline plots of the flow equations \eqref{eq:flow:nn_b} and \eqref{eq:flow:nn_u} for (a)~$\eps=0.1$ ($d<2$) and (b)~$\eps=-0.1$ ($d>2$). Closed symbols indicate the corresponding fixed points and the green open symbols indicate the initial points of a few trajectories (dotted lines). The eigenvectors of the fixed points are the straight lines denoted by the same color as the fixed point. While B$_-$ is repulsive, B$_+$ has one attractive and one repulsive direction. The separatrix is shown as the green dashed line: trajectories starting above the separatrix flow into the attractive fixed point (A for $d<2$ and G for $d>2$) while trajectories starting below run off to infinity.}
  \label{fig:neural}
\end{figure*}

As a second application, we consider AMB+~[Eq.~\eqref{eq:phi:ambp}] with $c_n=\kappa=0$,
\begin{equation}
  \partial_t\phi = \nabla^2(a \phi + b \phi^2 + u \phi^3) + \eta,
  \label{eq:neural}
\end{equation}
but with a noise term $\eta$ that is no longer conserved [$\al=0$ in Eq.~\eqref{eq:K}]. Equation~\eqref{eq:neural} has been derived recently for the evolution of a neural network~\cite{tiberi22}, where now $\phi(\x,t)$ is interpreted as a neural activity field, and is related to the famous Wilson-Cowan model~\cite{wilson72a}.

Since $\kappa=0$, we redefine the dimensionless non-linear model parameters with respect to $a$,
\begin{equation}
  \bar b^2 \equiv \frac{b^2D}{a^3}K_d\Lam^{d-2}, \qquad \bar u \equiv \frac{uD}{a^2}K_d\Lam^{d-2}.
\end{equation}
The graphical corrections~\cite{sm}
\begin{gather}
  \psi_a = 3 \bar u - 2 \bar b^2, \\
  \psi_b = 4\bar b^2 - 9\bar u, \\
  \psi_u = -9\bar u + 21\bar b^2 - 5\bar b^4/\bar u
\end{gather}
are calculated from all one-loop graphs shown in Fig.~\ref{fig:graphs}. Note that this model is different from the previous examples in that the expansion of the external wave vector is only up to second order. While graphical corrections to $\kappa$ and $c_{1,2}$ are generated, due to this $q$-expansion they are neglected. Turning to the graph in Fig.~\ref{fig:graph}(e), the graphical correction for the non-conserved noise term is $\mathcal{O}(q^2)$, which immediately implies $\psi_D=0$.

The final flow equations read
\begin{gather}
  \label{eq:flow:nn_b}
  \partial_\ell\bar b^2 = (2-d+14\bar b^2-27\bar u)\bar b^2, \\
  \label{eq:flow:nn_u}
  \partial_\ell\bar u = (2-d - 15\bar u+25\bar b^2)\bar u - 5\bar b^4
\end{gather}
for the two dimensionless non-linear parameters. In Fig.~\ref{fig:neural}, we plot the flow for two values of $\eps=2-d$ in the plane $(\bar b^2,\bar u)$. The Gaussian fixed point G at $(0,0)$ is attractive for $d>2$ and becomes repulsive for $d<2$. For $\bar b^2=0$, we find another perturbative fixed point A at $(0,\eps/15)$, which is attractive for $d<2$ and now governs the flow in the vicinity of G. For $d>2$, A becomes repulsive and we now find two more perturbative fixed points B$_\pm$ from the quadratic equation [after eliminating $\bar b^2$ by setting Eq.~\eqref{eq:flow:nn_b} to zero]
\begin{equation}
  \frac{2865}{196}\bar u^2 + \frac{29}{49}\eps\bar u - \frac{5}{196}\eps^2 = 0
\end{equation}
with solutions $\bar u^\ast=-\eps/15$ and $\bar u^\ast=5\eps/191$, and thus B$_+$ at $(-0.021\eps,0.026\eps)$ and B$_-$ at $(-0.200\eps,-0.067\eps)$. These fixed points require $\eps\leqslant 0$ so that $\bar b^2\geqslant 0$.

Starting from an initial point in the plane, there are two possible behaviors: either the flow runs into an attractive fixed point (A for $d<2$ and G for $d>2$) or the flow runs off to infinity. This could be an artifact of the one-loop approximation (Ref.~\cite{tiberi22} finds numerical evidence for $d=2$ that this is indeed the case) or it indicates the existence of a strong-coupling fixed point that is not accessible in our perturbative approach. Both behaviors are delineated by a line, the \emph{separatrix}. To calculate the separatrices, we recast Eqs.~\eqref{eq:flow:nn_b} and \eqref{eq:flow:nn_u} into a single differential equation $\frac{d \bar b^2}{d \bar u}$, which can be solved through an appropriate change of variables and substitution. For $\eps>0$ it can be shown that the line $y=\frac{1}{3}x$ is the separatrix. For $\eps<0$, the separatrix was found by integrating the flow equations backward in time starting close to the saddle point B$_+$ and using linear analysis close to the unstable points B$_-$ and A.


\section{Conclusions}

Dynamic renormalization is a powerful tool for analyzing large-scale behavior in active field theories, providing an efficient alternative to complementary field-theoretic approaches. In this manuscript, we first offer a concise but self-contained introduction to dynamic renormalization with applications, and we provide an accompanied python package for automating diagrammatic calculations. With these tools at our disposal, we then focus on conservative scalar field theories by revisiting Active Model B+, which extends the paradigmatic Model B of phase coexistence through including extra terms that break detailed balance and that has been argued to apply to scalar (non-aligning) active matter such as active Brownian particles~\cite{tjhung18}.

In particular the critical behavior of active Brownian particles has been studied intensively both in computer simulations~\cite{siebert18,partridge19,maggi21} and using theoretical arguments~\cite{speck22,dittrich21}. While here we take a decisive step forward, our results do not yet resolve whether motility-induced phase separation of active Brownian particles belongs to the same universality class as Model B. The reason is two-fold: First, although explicit construction schemes have been proposed~\cite{bickmann20a,speck22,vrugt23}, the precise relations between the microscopic model parameters (pair potential, speed, etc.) and the effective parameters $\vec x$ of the scalar field theory have not been established (and tested) yet. It is thus unclear what sector of AMB+, if at all, represents active Brownian particles.

\begin{table}[b!]
  \begin{tabular}{r|c|c|c|c|c|c|c|c}
  \hline
  Model & $a$ & $\kappa$ & $u$ & $b$ & $c_1$ & $c_2$ & $c_3$ & Remarks \\
  \hline
  \hline
  Linear & ($\times$) & ($\times$) & & & & & & \\
  cKPZ & ($\times$) & $\times$ & & & & $\times$ & \\
  cKPZ+ & ($\times$) & $\times$ & & & & $\times$ & $\times$ & \\
  Section~\ref{sec:mod_ckpz} & ($\times$) & $\times$ & & & $\times$ & & & \\
  Model B & $\times$ & ($\times$) & $\times$ & & & & & \\
  \hline
  Active Model B & $\times$ & $\times$ & $\times$ & $\times$ & $\times$ & $\times$ &  & $c_n \nabla^2 \phi^n$, $n\geq4$\\
  Active Model B+ & $\times$ & $\times$ & $\times$ & $\times$ & $\times$ & $\times$ & $\times$ & terms generated\\
  \hline
  \end{tabular}
  \caption{Overview of the field theories considered in this work that are contained within AMB+ [Eq.~\eqref{eq:phi:ambp}], where $\times$ denotes the non-zero model parameters. Parenthesis indicate parameters that could be set to zero and the resulting model would still be closed with respect to higher-order terms.}
  \label{table:models}
\end{table}

Second, as shown in Sec.~\ref{sec:higher}, AMB+ involves a truncation at order $\phi^3$ that lacks closure under one-loop dynamic renormalization since potentially relevant higher order non-linear terms arise at larger scales. Thence, we focus on the vicinity of the Wilson-Fisher and Gaussian fixed points, where these terms are small and can be neglected. We conclude that the Wilson-Fisher is the only non-trivial fixed point, and, therefore, that it determines the large-scale physics close to two dimensions. To explore a wider area of the parameter space, promising directions are two-loop renormalization~\cite{frey94} and functional renormalization approaches~\cite{canet11,jentsch23}. Finally, we demonstrate the efficiency of dynamic renormalization by applying it to the perturbatively accessible field theories derived from AMB+, see Table~\ref{table:models}. All results obtained for the examples considered in this work are equivalent to alternative action-based approaches and show how efficiently even complex flow equations can be obtained within dynamic renormalization. To tackle the integrals that arise from the vertex function Eq.~\eqref{eq:int:v2}, we have developed a python package that we hope will be particularly useful for scalar field theories involving similarly complicated vertex functions.


\begin{acknowledgements}
  We thank N. Goldenfeld for insightful discussions. This work has been supported by the Deutsche Forschungsgemeinschaft through TRR 146 (grant no. 233630050) and SFB 1551 (grant no. 464588647).
\end{acknowledgements}


\appendix

\section{Multiplicity of one-loop graphs}
\label{sec:mult}

\begin{figure}[b!]
  \centering
  \includegraphics{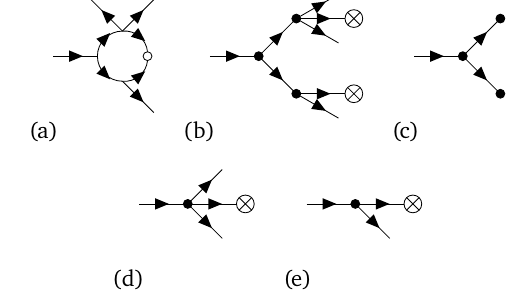}
  \caption{Example to illustrate the steps to calculate the multiplicity. (a)~Original graph [same as Fig.~\ref{fig:graphs}(j)]. (b)~The corresponding tree after cutting the open dot. (c-e)~The three vertices.}
  \label{fig:mult}
\end{figure}

To calculate the multiplicity of a given one-loop graph $\Gam$, we cut the correlation function and arrange the vertices into a tree. Each edge ends in either: another vertex (black dot), a noise term (crossed dot), or nothing. For each vertex $v_i$, the number of possible permutations of these three symbols is
\begin{equation}
  N_i = \frac{E!}{B!C!(E-B-C)!}
  \label{eq:perm}
\end{equation}
with $E$ the number of outgoing edges, $B$ the number of black dots, and $C$ the number of crossed dots. The graph multiplicity is $|\Gam|=\prod_i N_i$.

As an example, let us consider the graph Fig.~\ref{fig:mult}(a), which has three vertices in total. Its tree is shown in Fig.~\ref{fig:mult}(b). In Fig.~\ref{fig:mult}(c-e), we have isolated the vertices and their edges. Fig.~\ref{fig:mult}(c) has $E=2$, $B=2$, and $C=0$ with $N_1=1$. Fig.~\ref{fig:mult}(d) has $E=3$, $C=1$, and $B=0$ with $N_2=3$. Fig.~\ref{fig:mult}(e) has $E=2$, $C=1$, and $B=0$ with $N_3=2$. Multiplying the number of permutations, we obtain $|\Gam|=6$.

\section{Reconstructing vertex functions}
\label{sec:renorm_parameters}

Given the model parameters $\vec x$, let us assume that we have calculated the sum $I(\vec p_1,\dots,\vec p_n;\vec x)=\sum_m\mathcal I(\Gam^{m})$ of the one-loop graph integrals. We focus on a 2-vertex $v_2(\vec q,\vec p;\vec x)$ with scalar product $\vec q\cdot\vec p=qp\cos\psi$ but the following scheme extends to higher vertices. As already mentioned, we assume that $v_2=\sum_i x_iv_2^{(i)}$ is linear in the model parameters. Since vertex functions are polynomials, we expand $v_2^{(i)}=\sum_k a^{(i)}_kf_k$ in the monomial basis $f_k=\{q^2,p^2,qp\cos\psi,\dots\}$ with constant coefficients $a^{(i)}_k$. We expand the integral $I=\sum_kb_k(\vec x)f_k\delta\ell$ in the same basis. Using the definition of $\tilde x_i$ yields
\begin{multline}
  \tilde v_2 = \sum_i \tilde x_iv_2^{(i)} = \sum_i(1+\psi_{x_i}\delta\ell)x_iv_2^{(i)} \\ = v_2 + \sum_i\psi_{x_i}x_i \sum_k a^{(i)}_k f_k\delta\ell \overset{!}{=} v_2 + I
\end{multline}
and thus by comparing the basis coefficients
\begin{equation}
  \sum_i a^{(i)}_k x_i\psi_{x_i} = b_k(\vec x)
\end{equation}
for all $k$. This is a linear system of equations for the graphical corrections $\psi_{x_i}$ determined by the $b_k$ and the vertex structure encoded in the coefficients $a^{(i)}_k$.

\section{Flow of Model B in the presence of a non-zero cubic coefficient}
\label{sec:modelb_un}

\begin{figure*}[t]
  \centering
  \includegraphics{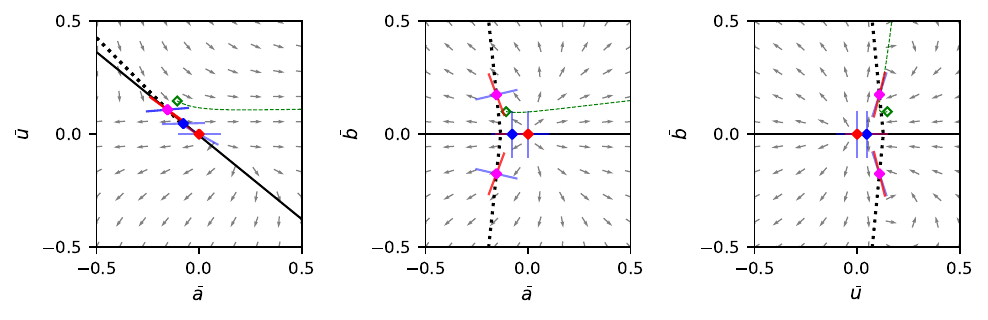}
  \caption{Renormalization flow portrait of Model B without shifting the origin of the field in $d=3.5$. The filled symbols are the fixed points: Gaussian (red), WF (blue) and C$_\pm$ (magenta). The arrows are the vector field centered on the Gaussian fixed point. The red and blue segments are the stable and unstable eigenvectors of each fixed point. The dotted lines are projections of the separatrix on the planes distinguishing the areas where $\bar{a} \rightarrow \pm \infty$.}
  \label{fig:model_b_separatrix}
\end{figure*}

Here we study Model B with evolution equation
\begin{equation}
  \partial_t \phi = \nabla^2 (a \phi+b \phi^2+u\phi^3-\kappa \nabla^2 \phi) + \eta
  \label{eq:phi_model_b}
\end{equation}
obtained from Eq.~\eqref{eq:phi:ambp} setting $c_1=c_2=c_3=0$. We apply the dynamic renormalization procedure to the reduced dimensionless parameters $\bar a$, $\bar u$ and $\bar b$ defined in \cref{eq:aubar,eq:bbar} obeying the flow equations
\begin{gather}
  \partial_l \bar{a} = \left(2+\psi_a -\psi_\kappa\right) \bar{a},
  \\
  \partial_l \bar{u} = \left(4-d+\psi_u-2\psi_\kappa\right) \bar{u},
  \\
  \label{eq:flow:b_modelb}
  \partial_l \bar{b} = \left(\frac{6-d}{2} -\frac{3}{2}\psi_\kappa +\psi_b\right) \bar{b},
\end{gather}
where the graphical corrections
\begin{gather}
  \psi_\kappa = \frac{\bar b^{2} \left(- \bar a^{2} d + 2 \bar a^{2} - \bar a d + 8 \bar a + 4\right)}{d(\bar a+1)^2},
  \\
  \psi_a = - \frac{2 \bar b^{2}}{\bar a (\bar a +1)^2} + \frac{3 \bar u}{\bar a (\bar a +1)},
  \\
  \psi_u = - \frac{5 \bar b^{4}}{\bar u (\bar a +1)^4} + \frac{21 \bar b^{2}}{(\bar a +1)^3} - \frac{9 \bar u}{(\bar a +1)^2},
  \\
  \psi_b = \frac{4 \bar b^{2}}{(\bar a +1)^3} - \frac{9 \bar{u}}{(\bar a +1)^2}
\end{gather}
are calculated from the one-loop graphs in Fig.~\ref{fig:graphs}~\cite{sm}. Note that from Eq.~\eqref{eq:flow:b_modelb}, for initially zero value, $\bar b$ stays zero and we recover Model B from Sec.~\ref{sec:modelb}. From these flow equations, we obtain the WF located at $\bar{a}^\ast=-\eps/6+\mathcal{O}(\eps^2), \bar{u}^\ast=\eps/9+\mathcal{O}(\eps^2)$ and $\bar{b}^\ast=0$ with $\eps=4-d$. Moreover, there are also new fixed points with $\bar b\neq0$ whose $\bar a$ values at any order of $\eps$ are given by the solutions of the polynomial
\begin{equation}
  \begin{split}
    \bar{a}^{6} \left(4 \eps^{2} - 16 \eps + 16\right) + \bar{a}^{5} \left(- 18 \eps^{2} - 248 \eps + 568\right) \\ + \bar{a}^{4} \left(- 54 \eps^{2} - 194 \eps + 1228\right) + \bar{a}^{3} \left(6 \eps^{2} + 688 \eps + 1668\right) \\ + \bar{a}^{2} \left(105 \eps^{2} + 1102 \eps + 1496\right) \\ + \bar{a} \left(90 \eps^{2} + 520 \eps + 536\right) + 23 \eps^{2} + 68 \eps + 44=0.
\end{split}
\end{equation}
We numerically study the solutions of this polynomial equation~\cite{sm} and we find that, for $\eps\in[0,2]$ ($d\in[2,4]$), there is only one new pair of fixed points C$_\pm$ with real model parameters. This pair is symmetric with respect to $\bar b \rightarrow - \bar b$, which originates from the symmetry $(\phi, b) \rightarrow (-\phi, -b)$ of Eq.~\eqref{eq:phi_model_b}. Importantly, we find that C$_\pm$ converges towards the WF for $\eps \rightarrow 2^+$ ($d\rightarrow 2^{-}$) and disappears for $\eps<0$ ($d>4$).

In Fig.~\ref{fig:model_b_separatrix}, we plot the flow in the three planes of the parameter space in $d=3.5$. We firstly revisit the stability of G and WF. For $\bar a$ and $\bar u$ we obtain the same linearized behavior as in Sec.~\ref{sec:modelb}. With our additional model parameter, we find that $\bar b$ is a relevant parameter for G if $d<4$ and for WF if $d\in(2,4)$. Note that $d=2$ is now a lower critical dimension for WF since its stability with respect to perturbations of $\bar b$ is different for $d>2$ and $d<2$. A heuristic argument for this critical dimension is that at exactly $d=2$ the fixed points C$_\pm$ joins with WF, thus changing its stability.

We now discuss the stability of the pair C of fixed points. Interestingly, each fixed point of C exhibits a conjugate pair of complex eigenvalues with negative real part. The corresponding eigenvectors are not aligned with a particular model parameter and have complex components in the plane ($\bar a$, $\bar u$). The complex nature of these eigenvalues leads to spiraling behavior of the flow close to C$_\pm$.

Additionally, there is a relevant direction that closely aligns with the $\bar{a}$ axis. We numerically study the flow~\cite{sm} and find that for non-zero $\bar{b}$ it follows this relevant direction converging towards trivial fixed points with $\bar a\rightarrow \pm \infty$ (and $\bar u$, $\bar b$ approximately constant) (see trajectory in Fig.~\ref{fig:model_b_separatrix} as an example). The only non-trivial critical point that the flow can converge to is the WF for $\bar b=0$ and $\bar a$, $\bar u$ placed on the line that connects G and WF in Fig.~\ref{fig:flow}. This observation bridges the analysis of Model B with the one obtained by shifting the origin of the field explained in Sec.~\ref{sec:modelb}.

\section{Graphical corrections for AMB+}
\label{sec:graphs_ambp}

We calculate the graphical corrections of AMB+ [Eq.~\eqref{eq:phi:ambp}] up to one-loop. For completeness, the function $h(q)$ defining the bare propagator [Eq.~\eqref{eq:G0}], the 2-vertex [Eq.~\eqref{eq:v2}], and the 3-vertex functions read
\begin{gather}
  h(q) = q^2 (a+\kappa q^2), \\
  \begin{split}
    v_2(\vec p_1,\vec p_2| \vec q) = &-b\vec q^2 + \frac{c_1}{2}\vec q^4 + c_2 \vec p_1 \cdot \vec p_2 q^2 \\ &-\frac{c_3}{2}(\vec p_2^2 \vec q\cdot \vec p_1+\vec p_1^2 \vec q\cdot \vec p_2),  
  \end{split} \\
  v_3(\vec q) = -u q^2,
\end{gather}
respectively. All of the possible one-loop graphs are shown in Fig.~\ref{fig:graphs}. 
The full expressions for the graphical corrections are too long to show here, but we provide the notebook used to calculate them~\cite{sm}. To lowest order of the model parameters, the corrections simplify to
\begingroup
\allowdisplaybreaks
\begin{align}
  & \begin{aligned} 
  \psi_a & = \frac{\eps \left(- 2 \bar b \bar c_3 + \bar c_1 \bar c_3\right)}{4 \bar a} - 3 \bar u + \\ 
  & \frac{ - 2 \bar b^{2} + \bar b \bar c_1 - 2 \bar b \bar c_2 + \bar c_1 \bar c_2 + 3 \bar u}{\bar a}, \\
  \end{aligned} \\
  & \begin{aligned} 
   \psi_\kappa & = 2 \bar b^{2} + \bar b \bar c_1 + 2 \bar b \bar c_2 - \frac{3 \bar b \bar c_3}{2} - \frac{\bar c_1^{2}}{2} +\bar c_1 \bar c_2 -
   \\ 
   &  \frac{\bar c_1 \bar c_3}{4} + \bar c_2^{2} - 2 \bar c_2 \bar c_3 + \eps \left( \frac{3 \bar c_1 \bar c_3}{16} - \frac{\bar c_2^{2}}{2} - \frac{5 \bar c_3^{2}}{16}\right) +
   \\
   & \eps \left(- \bar b^{2} + \frac{\bar b \bar c_1}{2} - \bar b \bar c_2 + \frac{\bar b \bar c_3}{8}\right), 
   \end{aligned} \\
& \psi_u = - 9 \bar u, \\
& \begin{aligned}
 \psi_b & = - \frac{3 \bar c_3 \eps \bar u}{4 \bar b} + \frac{- 9 \bar b \bar u + 3 \bar c_1 \bar u - 3 \bar c_2 \bar u}{\bar b},
\end{aligned} \\
 & \begin{aligned} 
 \psi_1 & = \frac{\eps \left(48 \bar b \bar u - 12 \bar c_1 \bar u + 24 \bar c_2 \bar u - 3 \bar c_3 \bar u\right)}{8 \bar c_1} +
 \\ & \frac{- 24 \bar b \bar u - 6 \bar c_1 \bar u - 12 \bar c_2 \bar u + 9 \bar c_3 \bar u}{2 \bar c_1},
\end{aligned} \\
& \begin{aligned} 
 \psi_2 & = \frac{\eps \left(- 6 \bar b \bar u + 3 \bar c_1 \bar u - 3 \bar c_3 \bar u\right)}{2 \bar c_2} + \frac{6 \bar b \bar u - 3 \bar c_3 \bar u}{\bar c_2}, \\
\end{aligned} \\
& \psi_3 = 0,
\end{align}
\endgroup
for which it suffices to calculate the graphs in Fig.~\ref{fig:graphs}(a,b,e,f,g). Here, $\eps=d-2$ and the dimensionless parameters $\bar x$ are given by \cref{eq:aubar,eq:bbar,eq:cbar}. Note that at this order of the model parameters, the flow of $\bar a$ decouples from the flow of the other model parameters.


%

\end{document}